\documentclass[%
 reprint,
 amsmath,amssymb,
 aps,
]{revtex4-2}

\usepackage{graphicx}% Include figure files
\usepackage{dcolumn}% Align table columns on decimal point
\usepackage{bm}% bold math
\usepackage{xcolor}
\usepackage{amsmath}
\usepackage{multirow}
\usepackage[left]{lineno}
\begin{document}
%\linenumbers
\preprint{APS/123-QED}

\title{Robust direct laser acceleration of electrons with flying-focus laser pulses}% Force line breaks with \\
\author{Talia Meir$^{1,2}$}
\email{taliameir@mail.tau.ac.il}
\author{Kale Weichman$^{3}$}
\author{Alexey Arefiev$^{4}$}
\author{John P. Palastro$^{3}$}
\author{Ishay Pomerantz$^{1}$}

\affiliation{% 
 $^{1}$The School of Physics and Astronomy, Tel Aviv University, Tel Aviv, 69978, Israel\\
 $^{2}$The School of Electrical Engineering, Tel Aviv University, Tel Aviv, 69978, Israel\\
 $^{3}$Laboratory for Laser Energetics, University of Rochester, Rochester, NY 14623 1299, USA\\
 $^{4}$Department of Mechanical and Aerospace Engineering, University of California San Diego, La Jolla, CA, 92093, USA
}%ution.edu}

\date{\today}% It is always \today, today,

\begin{abstract}
Direct laser acceleration (DLA) offers a compact source of high-charge, energetic electrons for generating secondary radiation or neutrons. While DLA in high-density plasma optimizes the energy transfer from a laser pulse to electrons, it exacerbates nonlinear propagation effects, such as filamentation, that can disrupt the acceleration process. Here, we show that superluminal flying-focus pulses (FFPs) mitigate nonlinear propagation, thereby enhancing the number of high-energy electrons and resulting x-ray yield. Three-dimensional particle-in-cell simulations show that, compared to a Gaussian pulse of equal energy (1 J) and intensity $(2\times10^{20} \ \mathrm{W/cm^2})$, an FFP 
produces $80\times$ more electrons above 100 MeV, increases the electron cutoff energy by 20\%, triples the high-energy x-ray yield, and improves x-ray collimation. These results illustrate the ability of spatiotemporally structured laser pulses to provide additional control in the highly nonlinear, relativistic regime of laser-plasma interactions.

\end{abstract}

\maketitle

%\tableofcontents

\section{Introduction}

Laser-driven sources of energetic electrons hold promise for a wide range of applications and fundamental studies, including electron radiography \cite{Schumaker2013,zhang2017femtosecond,Wan2022,Bruhaug2023}, neutron \cite{Pomerantz2014a,Gunther2022,Cohen2024_neutrons} and radiation generation \cite{Kneip2008,Cikhardt2024,Meir2024,Babjak2025,Tangtartharakul_2025,Proceeding2025}, and measurements of strong-field quantum electrodynamical phenomena \cite{Cole2018,blackburn2020radiation,Gonoskov2021Charged}. The extreme intensities required for the highest energy sources inevitably involve plasma due to the material breakdown limitations of solid-state or gaseous media. Intense, ultrashort laser pulses propagating through plasma accelerate electrons either directly, in the electromagnetic fields of the pulse itself, or indirectly, in the quasi-static fields that it drives. The direct laser acceleration (DLA) mechanism harnesses both processes: the first to energize electrons, and the second to laterally confine them, which facilitates their energy gain through the first process \cite{Arefiev2015}. The resulting high-efficiency transfer of laser energy to high-charge electron bunches has made DLA an attractive source that complements other laser-based approaches \cite{Malka1997,Gahn1999,Shaw_2018,Hussein_2021,Cohen2024,Babjak2024PRL,Rosmej2025}.

As an intense laser pulse propagates through plasma, its leading edge expels electrons from its path, forming a plasma channel. A population of electrons, either from the background plasma or edges of the channel, are injected into pulse, where they gain energy through its transverse electric field. The transverse momenta of these electrons are then rotated into the longitudinal direction by the magnetic field of the pulse. The quasi-static electric and magnetic fields generated during the formation of the plasma channel enhance the energy gain by providing transverse confinement \cite{pukhov1999particle} and driving transverse (betatron) oscillations \cite{Khudik2016Scaling}. When the Doppler-shifted laser frequency experienced by an electron becomes comparable to its betatron frequency, its transverse velocity can remain nearly antiparallel to the transverse electric field of the pulse over many cycles. The electron then gains energy with each oscillation. 

%A key feature of direct laser acceleration is the generation of quasi-static electric and magnetic fields in the plasma channel formed by the laser pulse. These fields provide transverse confinement for the accelerated electrons and strongly influence their dynamics \cite{pukhov1999particle}. Although the energy transfer originates from the transverse laser electric field, the forward-directed motion of relativistic electrons arises because the laser magnetic field redirects the acquired transverse momentum along the propagation axis. The quasi-static channel fields further enhance the energy gain by driving transverse (betatron) oscillations \cite{Arefiev2015}. When the betatron frequency becomes comparable to the laser frequency at the electron’s location, the transverse velocity can remain antiparallel to the oscillating laser electric field over many cycles, allowing the electron energy to increase with each oscillation.

\begin{figure*}[t]
\includegraphics[width=17.2cm]{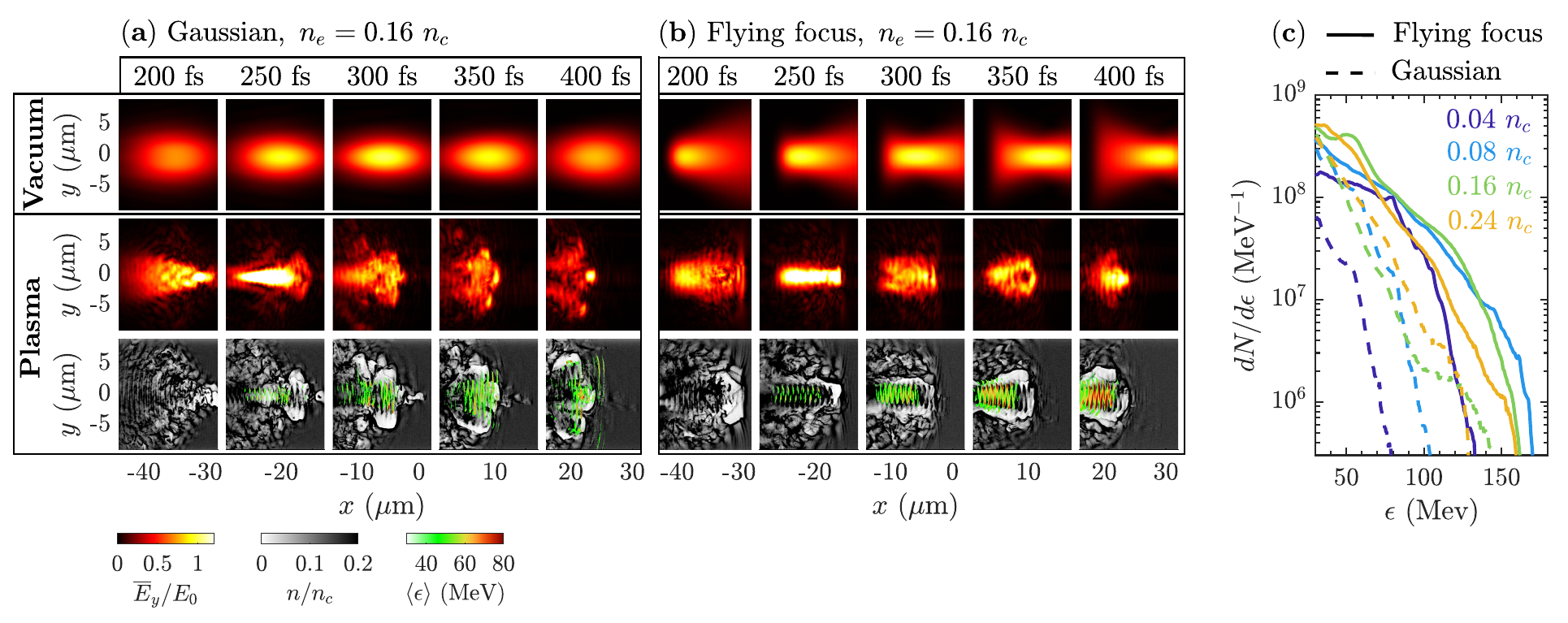}% Here is how to import EPS art
\caption{\label{Fig:FF_vs_G} Comparison of DLA with flying focus pulses (FFPs) and conventional Gaussian pulses (GPs) of equal energy and intensity. (Top row) (a) A GP propagates through vacuum and comes to focus at a fixed point in space. (b) A FFP propagates through vacuum with a focal point that moves at $v_f=1.2 c$. (Bottom rows) (a) The GP breaks up into multiple filaments, producing multiple small channels. (b) The FFP resists filamentation, producing a single, wide channel. (c) The wider and more uniform channel created by the flying focus results in more high energy electrons and a larger cutoff energy, independent of the electron density. Here, $\bar{E}_y/E_0$ is the transverse electric field envelope of the laser pulses normalized to their maximum amplitude in vacuum $E_0$, $n$ is the electron density, and $\epsilon$ the electron energy. The rainbow colorbar in the bottom row shows the average energy of electrons with energy greater than 30 MeV.
%(bottom row) The multiple
%3D simulation results of DLA driven by (a) a GP and (b) a FFP with a focal velocity $v_f=1.2 c$. The top and middle rows compare the vacuum propagation Envelope of $\mathrm{E_y/E_0}$ in a vacuum (top row), in the plasma (second row). Electron density in gray and the cell-average electron energy of electrons with $\mathrm{\epsilon > 30~MeV}$ in color (bottom row). (c) Electron spectrum results from simulations with $n_e = 0.04~n_c$ (dark blue), 0.08$~n_c$ (light blue), 0.16$~n_c$ (green), 0.24$~n_c$ (yellow), for FFLP with $v=1.2~c$ (solid line) and Gaussian beam (dashed line). At the time frame when the total kinetic energy of the electron population reaches its peak value (t = 500 fs, 500 fs, 400 fs and 350 fs, respectively).
}
\end{figure*}

While DLA can occur in low density plasma, the highest efficiencies and highest charge bunches are produced in plasmas where the electron density is comparable to the critical density---the density at which the plasma becomes opaque \cite{Babjak2024PRL}. At these near-critical densities, self-focusing, filamentation, and hosing can destabilize pulse propagation \cite{Najmudin2003, Esarey2009, Cros2014}, disrupting the structure of the quasi-static fields that are essential for sustained energy gain \cite{Willingale_2013, Hussein_2021}. To mitigate such instabilities, approaches based on structured targets, such as pre-fabricated micro-channels \cite{Snyder2019} or channels created by precursor pulses \cite{Vranic_2018} have been explored. These approaches, however, introduce their own challenges for experimental implementation, motivating the search for alternatives.

%While the ability to propagate through dense plasma and accelerate electrons makes DLA attractive, it also introduces challenges because the laser propagation can become unstable. In particular, intense laser pulses are susceptible to strong self-focusing, filamentation, and hosing-like instabilities \cite{Cros2014, Esarey2009, Najmudin2003}. These effects disrupt the structure of the quasi-static fields that are essential for sustained energy gain \cite{Khudik2016, Willingale_2013}. To mitigate such instabilities, approaches such as pre-formed micro-channels \cite{Snyder2019} or channels created by precursor pulses \cite{Vranic_2018} have been explored. These methods, however, introduce their own challenges for experimental implementation, which motivates the search for alternative approaches.

Flying-focus pulses (FFPs) offer a structured-light approach to stabilizing laser pulse propagation. FFPs feature an intensity peak that travels at an arbitrary, controllable velocity over distances far exceeding the Rayleigh range \cite{Sainte-Marie2017,Froula2018,palastro2020dephasingless,Caizergues2020,li2024spatiotemporal}. These features have been experimentally demonstrated using chromatic focusing of chirped laser pulses \cite{Froula2018,jolly2020controlling,Kabacinski2023,fu2025steering} and axilens focusing of pulses prepared by a radial echelon or refractive doublet \cite{Pigeon:24,Liberman2024}. The controllable velocity and extended interaction lengths afforded by FFPs benefit several laser-plasma applications, including electron acceleration \cite{palastro2020dephasingless,Caizergues2020,Ramsey2020} and  radiation generation \cite{Ramsey2022,Ye2023,simpson2024spatiotemporal}. Of particular note for DLA, FFPs have stabilized the propagation of low-intensity laser pulses in the formation of low-density plasma channels by mitigating ionization refraction \cite{Palastro2018,turnbull2018ionization}. The open question, which we address here, is whether the controllable velocity of a high-intensity flying-focus pulse can be used to stabilize a high-density plasma channel required for sustained DLA.

In this work, we demonstrate that a flying-focus pulse with an appropriately chosen superluminal velocity can enhance direct laser acceleration. Three-dimensional particle-in-cell simulations show that an FFP generates $80\times$ more electrons above 100 MeV, increases the cutoff energy by 20\%, and triples the x-ray yield above 200 keV compared to a standard Gaussian pulse (GP) with the same energy (1 J) and peak intensity $(2\times10^{20} \ \mathrm{W/cm^2})$. These improvements are linked to the ability of a superluminal FFP to maintain stable propagation and, consequently, a stable quasi-static field structure over the full acceleration length. With a Gaussian pulse, every temporal slice has the same nominal focal point. As a result, the back of the pulse must propagate through plasma that has already been disturbed by the front. With a superluminal FFP, earlier temporal slices focus deeper into the plasma than later slices. Thus, the back of the pulse has already come in and out of focus before encountering plasma that has been disturbed by the front. This allows the high-intensity peak of the FFP to traverse relatively quiescent plasma.

\section{Robust Electron Acceleration}

Figure~\ref{Fig:FF_vs_G} demonstrates that FFPs mitigate nonlinear propagation and enhance DLA relative to conventional GPs with the same energy and intensity. The figure displays the results of 3D particle-in-cell (PIC) simulations, comparing the propagation, channel formation, and electron acceleration for high-intensity GPs and FFPs in initially uniform, high-density plasma. The initial electron densities are typical of DLA experiments and range from $n_e = 0.04 - 0.24~n_c$, where $n_c = \varepsilon_0 m_e \omega_0^2/e^2$ is the critical density for the laser frequency $\omega_0$. The pulse parameters are motivated by commercially available Ti:Sapphire laser systems. The parameters are listed in Table I, while the simulation details are presented in Methods. 

%Figure~\ref{Fig:FF_vs_G} displays the results of 3D PIC simulations, comparing the propagation, channel formation, and electron acceleration for a conventional Gaussian pulse (GP) and a FFP. The physical parameters can be be found in Table I and were motivated by commercially available Ti:Sapphire laser systems. The simulation details are presented in Methods. 

Before describing the nonlinear dynamics underlying enhanced DLA in an FFP, it is instructive to outline the differences in the vacuum propagation of a GP and FFP. The top row of Fig.~\ref{Fig:FF_vs_G} illustrates these differences. The GP propagates along the $x$ axis at the vacuum speed of light $c$ and has a stationary focal plane at $x = 0$. As a result, the peak intensity of the GP occurs at a single time ($t \approx 300$ fs) and at a single point in space ($x = y = z = 0$). The FFP also propagates along the $x$ axis at $c$ but has a dynamic focal plane that travels at a superluminal velocity $v_f = 1.2 c$. In this case, the peak intensity moves from the back of the pulse to the front of the pulse in a time $T \approx 2v_f \tau /(v_f-c)$, where $\tau$ is the total pulse duration (see Table I). As a result, the peak intensity of the FFP occurs over a $T \approx 300$ fs window and a focal range $L_f = v_f T \approx 100$ $\mu$m.

Despite the strong currents excited by a high-intensity GP or FFP in high-density plasma, their linear propagation behavior still influences their nonlinear evolution. The middle row of Fig.~\ref{Fig:FF_vs_G} depicts this nonlinear evolution for $n_e = 0.16~n_c$. The GP becomes highly modulated and rapidly self focuses before breaking up into multiple filaments. The FFP, although also highly modulated, undergoes weaker self-focusing and resists filamentation. This improved stability of the FFP is due to the superluminal velocity of its intensity peak. The front of a superluminal FFP focuses deeper into the plasma than the back. As a consequence, the back of the pulse forms a high-intensity focus before having to propagate through plasma that has been disturbed by the front. Meanwhile, the front remains at lower intensity and traverses largely undisturbed plasma until it reaches its focus. The weaker plasma modifications experienced by the back and driven by the front mitigates self-focusing and filamentation, stabilizing propagation. This contrasts with a GP where the back of the pulse must propagate through plasma that has been strongly disturbed by the front, causing an accumulation of nonlinear refraction and focusing along the propagation path.

%undergoes strong self-focusing followed by filamentation, which depletes the field strength and disrupts the DLA structure. 

%An important finding of this work is that flying focus laser pulses overcome important limitations of conventional Gaussian pulses for driving direct laser acceleration in underdense plasmas close to the near-critical regime. As shown in Fig.~\ref{Fig:FF_vs_G}, FFLPs mitigate nonlinear propagation effects and enhance DLA relative to conventional Gaussian pulses with the same energy. FFLPs can produce higher-energy, higher-charge electron beams, and convert more laser energy into high-energy x-rays. Relative to conventional Gaussian pulses, FFLPs with the same energy are capable of producing higher-energy, higher-charge electron beams and converting more laser energy into high-energy X-rays. 

\begin{table}[b]
\caption{\label{tab:table_laser}%
Physical parameters for the target and pulses. The temporal electric field profile for both the GP and FFP was $\exp[-(t/\tau)^{2g}]$. Here, the normalized vector potential $a_0 = eE_0/m_e c \omega_0$.
}
\begin{ruledtabular}
\label{tab:sim-params}
\begin{tabular}{ll} %{lp{0.7cm}lp{2cm}}\multicolumn
{Target Parameters}&{Value}\\
\hline
Initial density $(n_{c})$&
{$0.04,~0.08,~0.16,~0.24$}\\
Density profile&
{Uniform, $\mathrm{-50~ \mu m<x<50~ \mu m}$}\\
Target dimensions 
 & {100$_x$ $\times$ 20$_y$ $\times$ 20$_z$ ($\mu$m$^3$)}\\
Initial composition &{Fully ionized carbon}\\
\end{tabular}
\begin{tabular}{lc}
{Shared Pulse Parameters}&{Value}\\
\hline
Wavelength $\lambda_0$ (nm)& 800 \\
Normalized vector potential $a_0$ & 9.8 \\
Intensity $(10^{20}$ W/cm$^2$) &2.04 \\
Energy (J)&1 \\
Polarization direction & $y$ \\
Duration $\tau$ (fs) &25 \\
\hline 
{Gaussian Pulse}&{} \\
\hline
Spot size $w_0$ (µm) &3.1 \\ %&2.6\\
Temporal profile order $g$ &1  \\
\hline 
{Flying Focus Pulse}&{} \\
\hline
Spot size $w_0$ (µm) &2.6 \\ 
Temporal profile order $g$ &4  \\
\end{tabular}
\end{ruledtabular}
\end{table}

%T(t), Temporal profile &\multicolumn{3}{l}{$\mathrm{\exp \left\{ -\left[ (t - t_0)^2/\tau^2 \right]^{g} \right\}}$} \\

A key component of DLA in high-density plasma is the formation of a channel, created when the leading edge of a laser pulse expels electrons leaving behind a positively charged column. The bottom row of Fig.~\ref{Fig:FF_vs_G} shows the density of the channel and the electrons accelerated within it. By propagating more stably, the FFP drives a wider, more uniform channel. This leads to a greater number of high-energy electrons and a larger cutoff energy regardless of the background density [Fig.~\ref{Fig:FF_vs_G}(c)]. For instance, at the optimal density of $0.08~n_c$, the FFP produces $80\times$ more electrons above 100 MeV than the GP, with a cutoff of 170 MeV compared to 140 MeV for the GP. 

\begin{figure}
\includegraphics[width=8.6cm]{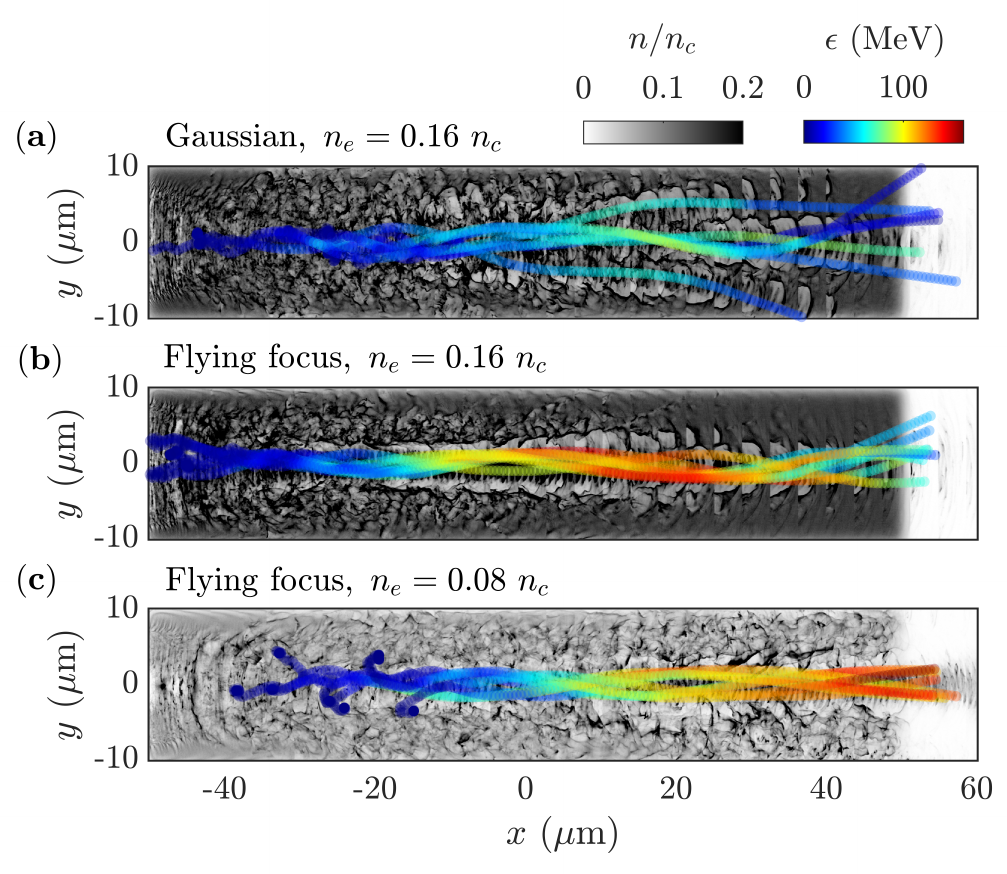}% Here is how to import EPS art
\caption{\label{fig:track} Enhanced collimation and energy gain of accelerated electrons in the wider, more-uniform channel produced by the FFP. The trajectories of representative high-energy electrons are colored by their energy at the corresponding location in the channel. The highest cutoff energy for a GP occurs for $n_e = 0.16~n_c$, whereas the highest cutoff for a FFP occurs for $n_e = 0.08~n_c$ [See Fig.~\ref{Fig:FF_vs_G}(c)]. The time-averaged density (gray) was calculated in a 8 $\mu$m window moving at the speed of light.}
%Trajectories of the most energetic electrons, For the Gaussian case, $\mathrm{n_e = 0.16~n_c}$ (a), and the FFLP case, v=1.2 c, with different densities: $\mathrm{n_e = 0.16~n_c}$ (b), $\mathrm{n_e = 0.08~n_c}$ (c). The energy along the trajectories (in color). The time-averaged density calculated using a 8 $\mu m$ window moving at the speed of light (gray).}
\end{figure}

The improved uniformity of the channel enhances DLA by increasing the collimation of the accelerated electrons. Figure 2 illustrates how the channel uniformity influences the trajectories and energy gain of representative high-energy electrons. In all cases, the electrons follow the path of the channel(s), whether its the multiple channels formed by the filamenting GP or the single, collimated channel formed by the FFPs. At the optimal density for the FFP [$n_e = 0.08~n_c$, Fig.~\ref{fig:track}(c)], the highest-energy electrons remain collimated and accelerate monotonically along the entire length of the channel. For visualization in Fig.~\ref{fig:track}, the electron density was averaged over snapshots taken every 25~fs, using an 8~$\mu$m-long window moving with the laser pulse (gray scale). The trajectories are shown for electrons randomly sampled from the most-energetic population, selected at the time where the energy spectrum had the largest cutoff (i.e., the right-most edge of the spectrum). 

The FFPs create wide, uniform channels, while filamentation of the GP results in multiple, smaller channels. The larger width of the channels produced by FFPs enhance the maximum energy gain achievable through DLA. The effect of channel width on the maximum energy gain can be understood using a simple model for the electron acceleration. Following the approach outlined in Ref.~\cite{Wang2020_PoP}, and as detailed in the Methods section, the maximum electron energy supported by a channel can be estimated as:
\begin{equation}
\gamma_{\text{max}} = \frac{\alpha}{\lambda_0^2}  \left(\frac{u}{u - 1}\right) R^2.
\label{eq:gamma_max}
\end{equation}
Here, \( R \) is the radius of the magnetic boundary defined as the maximum transverse displacement from the propagation axis where the magnetic field can still confine electrons to the channel, \(\alpha\) is proportional to the current, and \( u = v_\text{ph} / c \) is the normalized phase velocity of the laser pulse.
Equation~\ref{eq:gamma_max} indicates that the maximum energy increases when either the magnetic boundary expands, the current increases, or the phase velocity approaches unity. 

\begin{figure}[t]
\includegraphics[width=8cm]{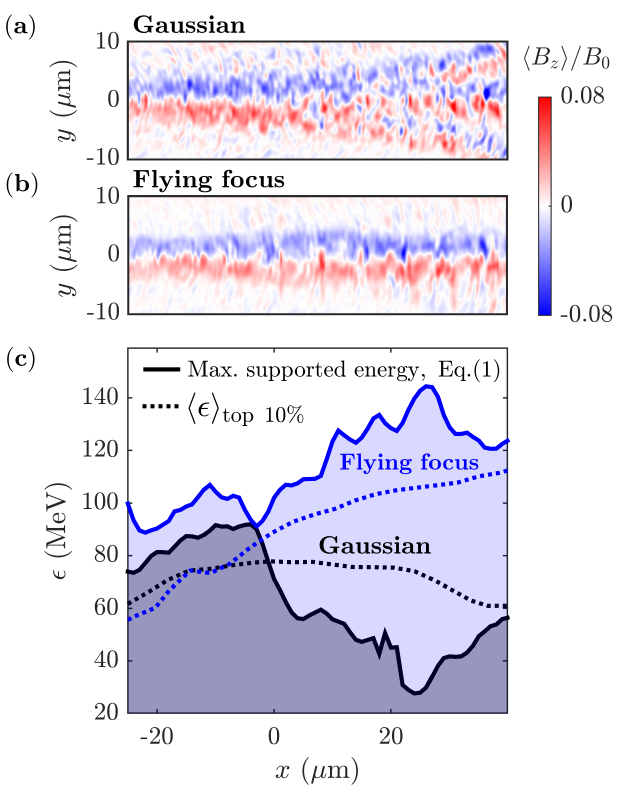}% Here is how to import EPS art
\caption{\label{fig:Bz_avg}
Magnetic field structure and maximum achievable electron energy in GP- and FFP-driven DLA. The channel produced by an FFP supports a higher maximum energy than that produced by a GP. (blue-to-red) The time-averaged azimuthal magnetic field driven by (a) the GP in a $n_e = 0.16~n_c$ plasma and (b) the $v_f = 1.2c$ FFP in a $n_e = 0.08~n_c$ plasma. Here, the magnetic field was averaged in a 30~$\mu$m window moving at the speed of light and normalized to the peak magnetic field of the laser pulse in vacuum $B_0$. The densities were chosen to maximize the cutoff energy for each pulse. (c) The maximum supported energy along the channel (solid curves) and range of energies up to the maxima (shaded) for the GP (black) and FFP (blue). The dashed lines show the average energy of the top 10\% most energetic electrons at each time step, plotted at their mean longitudinal position. The average energies increase monotonically until they approach the maximum energy supported by the channel. }
%Increase in the maximum supported DLA energy by an FFLP relative to a Gaussian pulse.
%The time-averaged azimuthal magnetic field (blue-to-red) for the Gaussian pulse, $\mathrm{n_e = 0.16~n_c}$  (a) and FFLP, $\mathrm{n_e = 0.08~n_c}$, with focal velocity $v$ = 1.2~c (b)
%calculated using a 30~$\mathrm{\mu m}$ window moving at the speed of light. 
%(c) The  maximal supported energy gain along the propagation axis (solid curves), with the range of energies up to this limit (shaded), for the Gaussian pulse (black) and FFLP (blue). The dashed line indicates the average energy of the top 10\% most energetic particles at each time step, plotted at their mean longitudinal position.}
\end{figure}

To demonstrate the effect of the channel width---and thus the magnetic boundary---on the electron energy, Eq.~\ref{eq:gamma_max} can be evaluated with $R$ and $\alpha$ values determined from the average magnetic field [Figs.~\ref{fig:Bz_avg}(a) and (b)] and a $u$ value obtained from the transverse field of the laser pulse (see Methods for details). As shown in Fig.~\ref{fig:Bz_avg}(c), DLA driven by an FFP supports higher maximum energies (solid curves) over a longer distance than DLA driven by a GP. In both cases, the average energy of the top 10\% most energetic electrons (dashed curves) increases  monotonically until it approaches the maximum supported by the channel. For the GP, the acceleration terminates near $x = 0$ where the pulse filaments; for the FFP the acceleration persists along the entire channel.

\begin{figure}
\includegraphics[width=8.6cm]{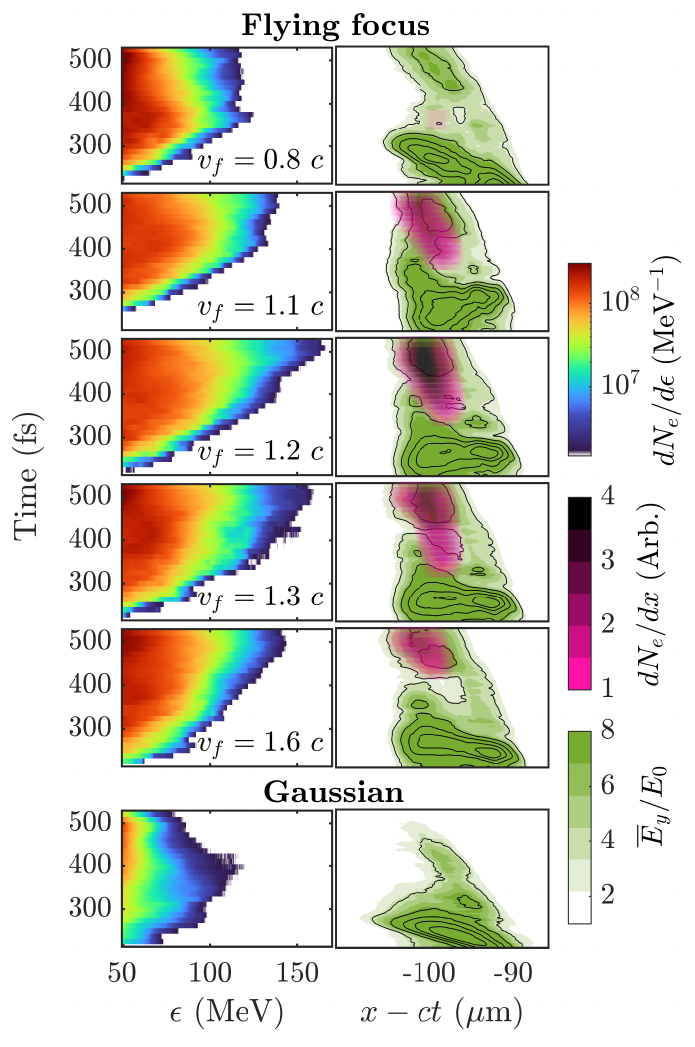}% Here is how to import EPS art
\caption{\label{fig:overlap} 
Optimization of the focal velocity $v_f$ for DLA and comparison of FFPs with the optimized GP. (left) Evolution of the electron energy spectrum. (right) Overlap between the transverse electric field envelope of the laser pulse (green) and electrons with energies greater than 100 MeV (pink). In all cases, the results for the optimal plasma density are displayed: $n_e = 0.08~n_c$ for the FFP and $n_e = 0.16~n_c$ for the GP. By stabilizing propagation, the superluminal FFPs allow for sustained overlap between the laser field and energetic electrons, leading to continued acceleration and higher energies.}
%Optimization of the FFP focal velocity for DLA and comparison with an optimized GP. The electron spectrum evolution with time (left), and The overlap of the field envelope with energetic electrons (right). The envelope of the transverse electric field (green) and the number of electrons with energy above 100 MeV (pink) evolve with time plotted in a moving window that moves with the speed of light. The figures are presented for an FFLP (top) for each focal velocity, $v/c = 0.8, 1.1,1.2,1.3,1.6$ (top to bottom) and for a Gaussian beam (bottom). Optimal plasma density chosen for each case, $\mathrm{n_e = 0.08~n_c}$ (FFLP) and  $\mathrm{n_e = 0.16~n_c}$ (Gaussian).}
\end{figure}

In addition to the channel width and uniformity, acceleration to high energies requires sustained overlap between the electrons and the high-intensity region of the laser pulse. For FFPs, this overlap is governed by the focal velocity. The optimal focal velocity $v_f = 1.2c$ presented in Figs.~\ref{Fig:FF_vs_G}--\ref{fig:Bz_avg} was identified by performing a simulation scan over $v_f$ values ranging from 0.8 -- 1.6 $c$. Figure~\ref{fig:overlap} shows the evolution of the electron energy spectrum (left column) and overlap between the laser pulse and energetic electrons (right column) for this range, with the GP results included at the bottom for comparison. The $v_f = 1.2c$ FFP sustains acceleration over the longest duration, leading to the highest electron energies. In all superluminal cases ($v_f > c$), where the propagation is more stable, a substantial population of electrons with energies above 100 MeV (pink colorbar) remain coincident with the transverse electric field of the FFPs (green colorbar), facilitating continued acceleration. This contrasts with the subluminal FFP ($v_f = 0.8 c$) and GP cases where the propagation is unstable and there are almost no electrons above 100 MeV. For fair comparison, each case is shown for the optimal initial electron density: $n_e = 0.08~n_c$ for the FFPs and $n_e = 0.16~n_c$ for the GP.

\section{Enhanced x-ray emission}

\begin{figure}[t]
\includegraphics[width=8.6cm]{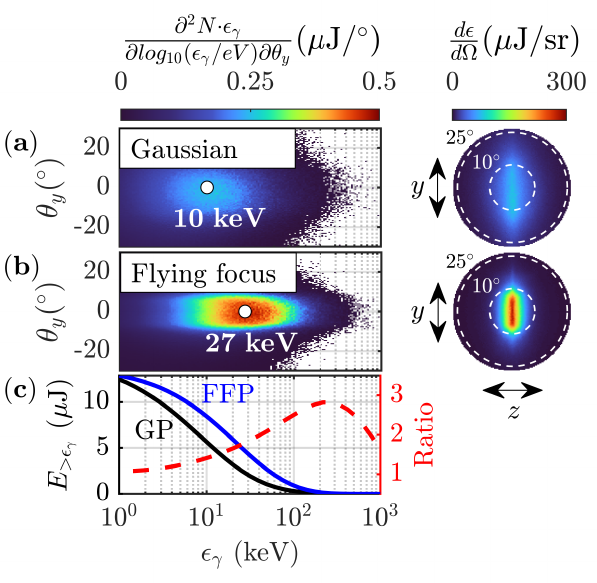}% Here is how to import EPS art
\caption{\label{fig:photons_theta_phi}
Enhancement of x-ray emission in FFP-driven DLA relative to GP-driven DLA. (left) The angularly resolved energy spectra of emitted photons for (a) the GP and (b) the $v_f = 1.2c$ FFP, each at its optimal electron density: $n_e = 0.16~n_c$ and $n_e = 0.08~n_c$, respectively. Here, $\theta_y$ is the emission angle of the photons in the polarization plane of the pulses, $y$-$z$. (right) Projections of the photon energy distributions on a unit sphere. The dashed circles indicate polar angles, and the arrows mark the $y$-axis (polarization direction) and $z$-axis. (c) Cumulative photon energy above $\mathrm{\epsilon_\gamma}$ for the GP (black) and FFP (blue). The dashed curve shows their ratio, demonstrating a threefold increase in conversion efficiency for photons above $\mathrm{200~keV}$ with the FFP.}
\end{figure}

The highly collimated, higher-energy electrons generated by the FFP emit x-rays that exhibit superior collimation and higher energies compared to those produced by the GP. This is demonstrated by the angle-resolved photon spectra presented in Figs. \ref{fig:photons_theta_phi} (a) and (b), where $\theta_y$ is the emission angle in the polarization plane ($y$-$x$) of the pulses. The spectral peak occurs at 27 keV for the FFP, compared to 10 keV for the GP. The angular distribution of the radiated energy is shown to the right, with the dashed lines marking the polar angle. For the FFP, the emission is more collimated and confined to a $10^{\circ}$ cone. Figure~\ref{fig:photons_theta_phi}(c) compares the cumulative energy of emitted photons above $\epsilon_\gamma$. The dashed curve shows the ratio of cumulative energies for the FFP and GP, demonstrating a three-fold enhancement in conversion efficiency for photons above $\mathrm{200~keV}$ with the FFP.

The enhanced x-ray emission with the FFP arises from the motion of high-energy electrons in the electromagnetic fields of the laser pulse and quasi-static fields of the plasma. During DLA, the transverse (i.e., betatron) oscillations of electrons in the combined laser and plasma fields generate x-rays through two distinct mechanisms \cite{Tangtartharakul_2025}. In the first mechanism, the rapid acceleration within each cycle of the laser pulse generates x-rays that are emitted in a double-lobed angular structure about the propagation axis. Observation of this mechanism is indicative of relatively inefficient electron acceleration, characterized by frequent cycles of energy gain and loss. In the second, more efficient mechanism, electrons gradually gain energy over multiple cycles, interact with the azimuthal magnetic field of the plasma, and emit x-rays in a single-lobed, collimated angular pattern. 

\begin{figure}[t]
\includegraphics[width=8.6cm]{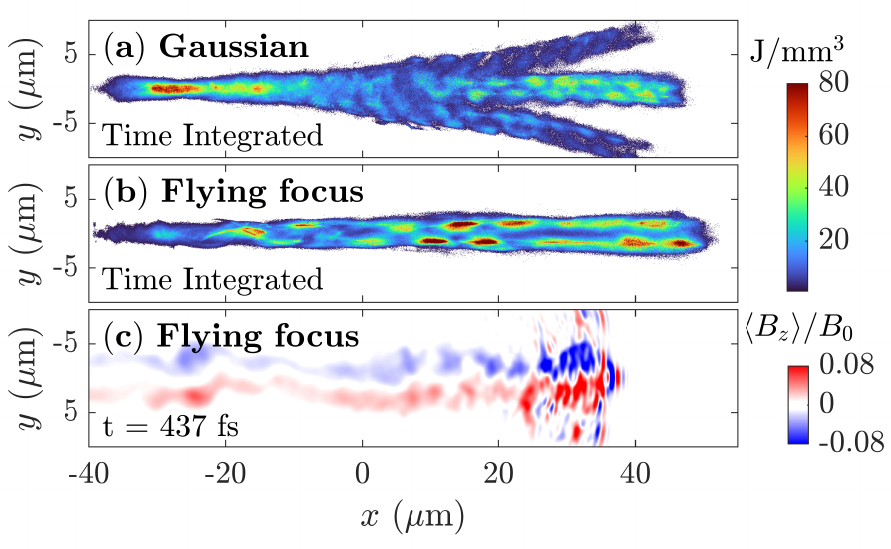}% Here is how to import EPS art
\caption{\label{fig:photons_xy} Location of photon emission in GP- and FFP-driven DLA. The GP results in relatively inefficient DLA, characterized by photon emission at the center of the channel. The FFP, in contrast, results in efficient DLA and emission at the channel edges, coincident with quasi-static magnetic islands. (a,b) Time-averaged energy density of $\mathrm{\epsilon_{\gamma} > 1~keV}$  photons generated at each location for (a) the GP and (b) the $v_f = 1.2c$ FFP, each at its optimal density: $n_e = 0.16~n_c$ and $n_e = 0.08~n_c$, respectively. (c) Magnetic field for the FFP case at $t$ = 437 fs and $z$ = 0, spatially averaged and normalized to the peak magnetic field of the laser pulse in vacuum $B_0$.}
\end{figure}

Figure~\ref{fig:photons_xy} displays the time-integrated energy density of x-rays generated at each location, revealing the origin of the enhanced emission. For the GP, the emission peaks near the center of the channel at $x \approx -25~\mu$m. This is a hallmark of the inefficient mechanism \cite{Tangtartharakul_2025},
where the emission is dominated by the fields of the laser pulse rather than the quasi-static magnetic field. For the FFP, the emission peaks at the edges of the channel in localized islands. Inspection of the quasi-static magnetic field at a representative time $t=\mathrm{437~fs}$ [Fig.~\ref{fig:photons_xy}(c)] shows a clear correlation between the emission and island-like magnetic-field structure. These quasi-periodic magnetic islands arise from current vortices formed in the wake of the laser pulse \cite{Gong2021,Cai_2025}. This structure is consistent with the efficient x-ray generation mechanism described above.

\section{Discussion}

FFPs feature an arbitrary, tunable velocity intensity peak, providing an additional degree of freedom for optimizing DLA. The 3D PIC simulations presented in this work demonstrate that a FFP with a superluminal ($v_f > c$) intensity peak mitigates the nonlinear propagation effects that otherwise limit the performance of DLA in high-density plasma. By stabilizing propagation, the FFP addresses two key requirements for efficient DLA: (1) formation of a stable, uniform channel and (2) sustained overlap between the laser pulse and high-energy electrons. 

The wider, more uniform channel and sustained overlap enabled by the FFP improve both the collimation and energy gain of accelerated electrons. Compared to an optimized GP with the same energy (1 J) and intensity $(2\times10^{20} \ \mathrm{W/cm^2})$, the FFP produces $80\times$ more electrons above 100 MeV and raises the cutoff energy by 20\%. These improvements in the channel structure and electron dynamics compound to generate x-rays with higher energies and reduced angular divergence: the FFP triples the yield of x-rays above 100 keV while simultaneously enhancing their collimation. For the optimal focal velocity ($v_f = 1.2c$) and density ($n_e = 0.08~n_c$), FFP-driven DLA generates approximately $\mathrm{10^{10}}$ photons with energy above $\mathrm{1~keV}$, 
with a total energy of $\mathrm{14.2~\mu J}$, corresponding to an IR-to-x-ray conversion efficiency of $\mathrm{\sim 10^{-5}}$. The resulting source brightness is approximately 
$\mathrm{10^{21}}$~(s~mrad$^2~$mm$^2~$0.1$\%$BW)$^{-1}$,
comparable to measurements of DLA-generated x-rays using GPs with 20 times higher energy \cite{Rosmej2021,Cikhardt2024}.

The enhanced stability of high-intensity FFPs demonstrated here suggests that FFPs may benefit other laser-plasma applications that rely on the formation of a uniform channel, such as ion acceleration in near-critical density gas targets \cite{Singh2020,Ospina-Bohrquez2024} or self-guiding of relativistically intense laser pulses~\cite{PhysRevResearch.2.043173}. Nonlinear propagation dynamics can also limit electron acceleration, x-ray generation, and electron-positron pair creation, in relativistically transparent ($n_e < a_0 n_c$) or ultra-relativistic regimes ($a_0 \gtrsim  100$) \cite{Wang2020}. Future work will consider the potential impact of FFPs in these regimes.

\section{Methods}
\label{sec:methods}
\subsection{Simulation setup}
\begin{table}[b]
\caption{\label{tab:table1}%
numerical parameters.
}
\begin{ruledtabular}
\label{tab:sim-params1}
\begin{tabular}{ll}{Numerical Parameters}\\
\hline
Macro-particles per cell
~& {12 e$^-$, 2 C$^{+6}$}\\
Simulation box  
 & {160$_x$ $\times$ 24$_y$ $\times$ 24$_z$ ($\mu$m$^3$)}\\
Spatial resolution 
 & {$\mathrm{25 \times 25 \times 25}$ (cells $\mu$m$^{-3}$)}\\
\end{tabular}
\end{ruledtabular}
\end{table}

The advantages of FFPs over conventional GPs were explored through a series of 3D simulations using the open-source, fully relativistic PIC code EPOCH \cite{Arber2015}. In all cases, the FFPs and GPs were initialized with the same energy, pulse duration, and normalized vector potential. This was achieved by making slight adjustments to the spot size and temporal profile of the pulses. The pulse parameters are listed in Table~\ref{tab:table_laser}, while the target and numerical parameters are provided in Table~\ref{tab:table1}.

%We used the fully relativistic particle-in-cell (PIC) code EPOCH \cite{Arber2015} in 3D,  The target parameters and resolution are summarized in Table~\ref{tab:table1}. 

The numerical implementation of FFPs follows Ref.~\cite{Franke2021}. The transverse electric field is initialized at a single location in $x$ with a Gaussian  profile and a time-dependent focal length: $f(t) = f_0 + c v_f t/(c-v_f)$, where $v_f $ is the focal velocity. The time-dependent focal length determines the time-dependent curvature 
\( R(t) = f(t)[1 + Z_R^2 / f^2(t)] \), 
spot size \( w(t) = w_0[1 + f^2(t) / Z_R^2]^{1/2} \), and 
Gouy phase \( \psi(t) = \tan^{-1}[f(t) / Z_R] \), where $w_0$ is the spot size at focus and \( Z_R = \pi w_0^2/\lambda_0 \) is the Rayleigh range.
The initial field is then 
\begin{equation}
\label{eq:FF_fields}
\begin{aligned}
E(r,t) = & \, E_0 \frac{w_0}{w(t)} \exp \left( -\frac{r^2}{w^2(t)} \right) \\& \times \exp \left[ -i \left( \omega t - k \frac{r^2}{2R(t)} +\psi(t)\right) \right]  T(t),
\end{aligned}
\end{equation}
where $E_0$ is the maximum electric field in vacuum, $k=2\pi/\lambda_0$, $\omega = ck$, and $T(t)$ is a temporal profile. For the simulations presented here $T(t) = \exp[-(t/\tau)^{2g}]$ (see Table~\ref{tab:table_laser}).

\subsection{X-ray emission}
EPOCH features a Monte-Carlo based module for calculating the synchrotron emission \cite{Ridgers2014}, which depends on the effective transverse field strength:
\begin{align}
  H \equiv \sqrt{ \left( \mathbf{E} + \frac{1}{c} \mathbf{v} \times \mathbf{B} \right)^2 - \frac{1}{c^2} (\mathbf{v} \cdot \mathbf{E})^2 }
\label{eq:H_def}
\end{align}
where $\mathbf{E}$ and $\mathbf{B}$ are the electromagnetic fields at the location of an electron with velocity $\mathbf{v}$. In the classical limit relevant to this work, the module statistically reproduces the known expression for the radiated energy per unit frequency~\cite{landau_lifshitz_fields}:
\begin{equation}
\frac{dI}{dw} = \frac{\sqrt{3} e^3 H}{2 \pi m c^2} \frac{\omega}{\omega_c}  \int_{\omega/\omega_c}^{\infty} K_{5/3}(x) \, dx,
\end{equation}
where $\omega_c = 3eH\gamma^2/2mc$ is the critical synchrotron frequency,
$K_{n}$ is the modified Bessel function of the second kind, and $\gamma = (1-\mathbf{v}\cdot\mathbf{v}/c^2)^{-1/2}$ is the Lorentz factor of the radiating electron.

%, which depends on the effective transverse field strength:
%\begin{align}
%  H \equiv \sqrt{ \left( \mathbf{E} + \frac{1}{c} \mathbf{v} \times \mathbf{B} \right)^2 - \frac{1}{c^2} (\mathbf{v} \cdot \mathbf{E})^2 }
%  \label{eq:H_def}
%\end{align}
%where $\mathbf{E}$ and $\mathbf{B}$ are the electromagnetic fields at the location of an electron with velocity $\mathbf{v}$. 
%The code uses a full quantum description of synchrotron emission
%to calculate the radiated energy per unit frequency,
%which in the classical limit relevant to this work is reduced to~\cite{landau_lifshitz_fields}:
%\begin{equation}
%\frac{dI}{dw} = \frac{\sqrt{3} e^3 H}{2 \pi m c^2} \frac{\omega}{\omega_c}  \int_{\omega/\omega_c}^{\infty} K_{5/3}(x) \, dx
%\end{equation}
%where $\omega_c = 3eH\gamma_e ^2/2mc$ is the critical synchrotron frequency,
%$K_{n}$ is the modified Bessel function of the second kind, and $\gamma_e$ is the Lorentz factor of the radiating electron.
 
\subsection{Maximum electron energy supported by a DLA channel}

Following Ref.~\cite{Wang2020_PoP}, the fields acting on an electron are separated into the electromagnetic fields of the laser pulse and the quasi-static fields of the channel. The laser pulse is approximated as a plane wave with a superluminal phase velocity $v_{\text{ph}}$ that accounts for both the dispersive properties of the plasma and the finite transverse extent of the laser beam. 
%we take the standard approach of separating the forces acting on the electrons into a superposition of the laser and the channel fields. We approximate the laser as a plane electromagnetic wave with a superluminal phase velocity that accounts for both the dispersive properties of the plasma and the finite transverse extent of the laser beam. 
The problem then reduces to solving the relativistic equations of motion under the action of
\begin{equation}
\mathbf{E} = \mathbf{e}_y E_0 \cos(\xi), \quad
\mathbf{B} = \mathbf{e}_z \frac{c}{v_{\text{ph}}} E_0 \cos(\xi) + \mathbf{B}_{\text{ch}},
\end{equation}
where $\bold{e}_j$ denotes a unit vector in the direction $j$. The channel field $\mathbf{B}_{\text{ch}}$ is derived from its normalized vector potential
\begin{equation}
\mathbf{a}_{\text{ch}} = \mathbf{e}_x \, \alpha \frac{r^2}{\lambda_0^2}, \quad
\mathbf{B}_{\text{ch}} = \frac{m_e c^2}{|e|} \nabla \times \mathbf{a}_{\text{ch}}.
\end{equation}
where \( \lambda_0 \) is the laser wavelength and \( \alpha \) is defined as
\begin{equation}
\alpha = -\frac{2\pi \lambda_0^2 |e| j_0}{m_e c^3} = -\frac{|e| \lambda_0^2}{2 m_e c^2} \frac{\partial \langle B_z \rangle}{\partial y}.
\label{eq:alpha}
\end{equation}
Here, \( j_0 \) is the longitudinal current density 
expressed in terms of the magnetic field gradient through Ampere's law.

The fields and particle dynamics depend on a single coordinate $\xi =~ \omega_0(t - x/v_\text{ph})$, which leads to the conserved integral of motion:
\begin{equation}
\label{eq:conserved_Q}
\gamma - \frac{v_{\text{ph}}}{c} \frac{p_x}{m_e c} + \frac{v_{\text{ph}}}{c} a_{\text{ch}}
= \text{const.}
= \gamma_i,
\end{equation}
where $p_x$ is the longitudinal momentum of the electron and $\gamma_i$ the initial value of its Lorentz factor
before it enters the wave. Equation \ref{eq:conserved_Q} can be rearranged as
\begin{equation}
u a_{\text{ch}} = \gamma_i - \left[ \gamma - \frac{p_x}{m_e c} \right] + (u - 1) \frac{p_x}{m_e c},
\label{eq:conserved_Q2}
\end{equation}
where \( u = v_\text{ph} / c \) is the normalized phase velocity. As the longitudinal momentum and Lorentz factor \( \gamma \) increase, the condition \( \gamma - ~p_x/m_e c  > 0 \) must hold. Since \( u - 1 \geq 0 \), the maximal energy gain is reached when \( \gamma - ~p_x/m_e c  \to 0 \). In this limit, the first term in Eq.~\ref{eq:conserved_Q2} can be neglected
and the maximum energy gain approximated by \( p_x/m_e c ~ \approx \gamma_\text{max} \). For the regime of interest, \( \gamma / \gamma_i \gg (u - 1)^{-1} \) and 
$\gamma_\text{max}$ can be estimated by combining the above equations, yielding Eq. \ref{eq:gamma_max}.

\begin{figure}[t]
\includegraphics[width=8.6cm]{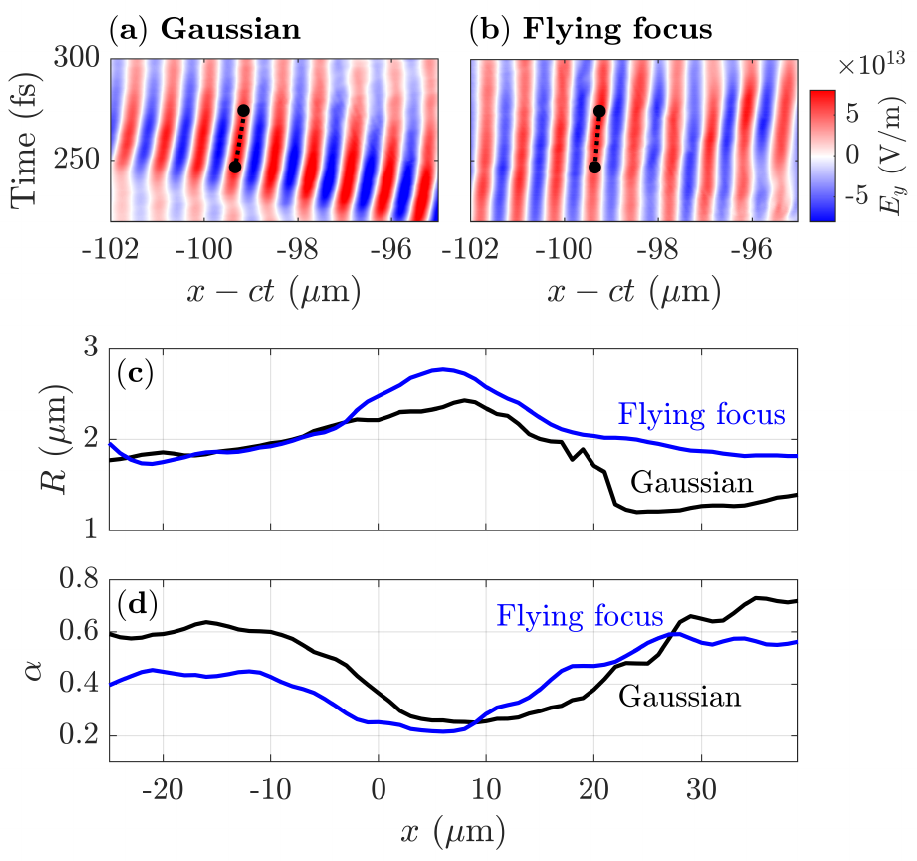}% Here is how to import EPS art
\caption{\label{fig:u_R_alpha}
Parameters used to estimate the maximum achievable energy. (a,b) On-axis ($y = z = 0$) temporal profiles of the transverse electric field for (a) the  GP and (b) the $v_f = 1.2c$ FFP, each at its optimal density: $n_e = 0.16~n_c$ and  $n_e = 0.08~n_c$, respectively. The black dashed lines show the segments used to find $v_\mathrm{ph}$: 1.0215~$c$ for the GP and 1.0119~$c$ for the FFP. (c) The average radius $R$ of the quasi-static magnetic field and (d) the parameter $\alpha$, which is proportional to the current density, are evaluated at each $x$-position for both the GP (black) and the FFP (blue).}
\end{figure}

To calculate the maximum supported energy in Fig.~\ref{fig:Bz_avg}~(c), the phase velocity was extracted from the transverse electric field plotted in Fig.~\ref{fig:u_R_alpha}.
The parameters $R$ and $\alpha$ were evaluated at each $x$-position using a 20~$\mu$m moving average window of the  magnetic field plotted in Fig.~\ref{fig:Bz_avg}. The parameter $\alpha$ was obtained from the slope $\partial \langle B_z \rangle / \partial y$, and $2R$ was defined as the distance between the minimum and maximum of $\langle B_z \rangle$ along the $y$-axis.

\section{Data availability}
The data sets analyzed during the current study are available from the corresponding author on reasonable request.

\section{Code availability}
All codes used during the current work are available from the corresponding
author upon reasonable request.

\section{Acknowledgments}
This research was supported by Grant No. 2022322 from the United States-Israel Binational Science Foundation (BSF). We acknowledge the EuroHPC Joint Undertaking for awarding this project access to the EuroHPC supercomputer LUMI, hosted by CSC (Finland) and the LUMI consortium through a EuroHPC Regular Access call. Simulations were performed using EPOCH, which was developed as part of the UK Engineering and Physical Sciences Research Council
(EPSRC) grants EP/G054950/1, EP/G056803/1, EP/G055165/1 and EP/ M022463/1.
T.M. acknowledges support by the Jabotinsky Fellowship of the Ministry of Science and Technology, Israel. The work of K.W. and J.P.P. was supported by the Department of Energy National Nuclear Security Administration under award No. DE-NA0004144 and the Department of Energy Office of Fusion Energy Sciences under award No. DE-SC0021057. 

\section{Author contributions}
T.M. performed and analyzed the PIC simulations. All authors wrote the manuscript. I.P. and A.A. obtained funding specific to this project.

\section{Competing interests}
The authors declare no competing interests.

%\nocite{*}

\bibliography{bib_ref}% Produces the bibliography via BibTeX.

\begin{thebibliography}{60}%
\makeatletter
\providecommand \@ifxundefined [1]{%
 \@ifx{#1\undefined}
}%
\providecommand \@ifnum [1]{%
 \ifnum #1\expandafter \@firstoftwo
 \else \expandafter \@secondoftwo
 \fi
}%
\providecommand \@ifx [1]{%
 \ifx #1\expandafter \@firstoftwo
 \else \expandafter \@secondoftwo
 \fi
}%
\providecommand \natexlab [1]{#1}%
\providecommand \enquote  [1]{``#1''}%
\providecommand \bibnamefont  [1]{#1}%
\providecommand \bibfnamefont [1]{#1}%
\providecommand \citenamefont [1]{#1}%
\providecommand \href@noop [0]{\@secondoftwo}%
\providecommand \href [0]{\begingroup \@sanitize@url \@href}%
\providecommand \@href[1]{\@@startlink{#1}\@@href}%
\providecommand \@@href[1]{\endgroup#1\@@endlink}%
\providecommand \@sanitize@url [0]{\catcode `\\12\catcode `\$12\catcode `\&12\catcode `\#12\catcode `\^12\catcode `\_12\catcode `\%12\relax}%
\providecommand \@@startlink[1]{}%
\providecommand \@@endlink[0]{}%
\providecommand \url  [0]{\begingroup\@sanitize@url \@url }%
\providecommand \@url [1]{\endgroup\@href {#1}{\urlprefix }}%
\providecommand \urlprefix  [0]{URL }%
\providecommand \Eprint [0]{\href }%
\providecommand \doibase [0]{https://doi.org/}%
\providecommand \selectlanguage [0]{\@gobble}%
\providecommand \bibinfo  [0]{\@secondoftwo}%
\providecommand \bibfield  [0]{\@secondoftwo}%
\providecommand \translation [1]{[#1]}%
\providecommand \BibitemOpen [0]{}%
\providecommand \bibitemStop [0]{}%
\providecommand \bibitemNoStop [0]{.\EOS\space}%
\providecommand \EOS [0]{\spacefactor3000\relax}%
\providecommand \BibitemShut  [1]{\csname bibitem#1\endcsname}%
\let\auto@bib@innerbib\@empty
%</preamble>
\bibitem [{\citenamefont {Schumaker}\ \emph {et~al.}(2013)\citenamefont {Schumaker}, \citenamefont {Nakanii}, \citenamefont {McGuffey}, \citenamefont {Zulick}, \citenamefont {Chyvkov}, \citenamefont {Dollar}, \citenamefont {Habara}, \citenamefont {Kalintchenko}, \citenamefont {Maksimchuk}, \citenamefont {Tanaka}, \citenamefont {Thomas}, \citenamefont {Yanovsky},\ and\ \citenamefont {Krushelnick}}]{Schumaker2013}%
  \BibitemOpen
  \bibfield  {author} {\bibinfo {author} {\bibfnamefont {W.}~\bibnamefont {Schumaker}}, \bibinfo {author} {\bibfnamefont {N.}~\bibnamefont {Nakanii}}, \bibinfo {author} {\bibfnamefont {C.}~\bibnamefont {McGuffey}}, \bibinfo {author} {\bibfnamefont {C.}~\bibnamefont {Zulick}}, \bibinfo {author} {\bibfnamefont {V.}~\bibnamefont {Chyvkov}}, \bibinfo {author} {\bibfnamefont {F.}~\bibnamefont {Dollar}}, \bibinfo {author} {\bibfnamefont {H.}~\bibnamefont {Habara}}, \bibinfo {author} {\bibfnamefont {G.}~\bibnamefont {Kalintchenko}}, \bibinfo {author} {\bibfnamefont {A.}~\bibnamefont {Maksimchuk}}, \bibinfo {author} {\bibfnamefont {K.~A.}\ \bibnamefont {Tanaka}}, \bibinfo {author} {\bibfnamefont {A.~G.~R.}\ \bibnamefont {Thomas}}, \bibinfo {author} {\bibfnamefont {V.}~\bibnamefont {Yanovsky}},\ and\ \bibinfo {author} {\bibfnamefont {K.}~\bibnamefont {Krushelnick}},\ }\bibfield  {title} {\bibinfo {title} {Ultrafast electron radiography of magnetic fields in high-intensity laser-solid interactions},\ }\href
  {https://doi.org/10.1103/PhysRevLett.110.015003} {\bibfield  {journal} {\bibinfo  {journal} {Phys. Rev. Lett.}\ }\textbf {\bibinfo {volume} {110}},\ \bibinfo {pages} {015003} (\bibinfo {year} {2013})}\BibitemShut {NoStop}%
\bibitem [{\citenamefont {Zhang}\ \emph {et~al.}(2017)\citenamefont {Zhang}, \citenamefont {Hua}, \citenamefont {Wan}, \citenamefont {Pai}, \citenamefont {Guo}, \citenamefont {Zhang}, \citenamefont {Ma}, \citenamefont {Li}, \citenamefont {Wu}, \citenamefont {Chu} \emph {et~al.}}]{zhang2017femtosecond}%
  \BibitemOpen
  \bibfield  {author} {\bibinfo {author} {\bibfnamefont {C.}~\bibnamefont {Zhang}}, \bibinfo {author} {\bibfnamefont {J.}~\bibnamefont {Hua}}, \bibinfo {author} {\bibfnamefont {Y.}~\bibnamefont {Wan}}, \bibinfo {author} {\bibfnamefont {C.-H.}\ \bibnamefont {Pai}}, \bibinfo {author} {\bibfnamefont {B.}~\bibnamefont {Guo}}, \bibinfo {author} {\bibfnamefont {J.}~\bibnamefont {Zhang}}, \bibinfo {author} {\bibfnamefont {Y.}~\bibnamefont {Ma}}, \bibinfo {author} {\bibfnamefont {F.}~\bibnamefont {Li}}, \bibinfo {author} {\bibfnamefont {Y.}~\bibnamefont {Wu}}, \bibinfo {author} {\bibfnamefont {H.-H.}\ \bibnamefont {Chu}}, \emph {et~al.},\ }\bibfield  {title} {\bibinfo {title} {Femtosecond probing of plasma wakefields and observation of the plasma wake reversal using a relativistic electron bunch},\ }\href@noop {} {\bibfield  {journal} {\bibinfo  {journal} {Physical review letters}\ }\textbf {\bibinfo {volume} {119}},\ \bibinfo {pages} {064801} (\bibinfo {year} {2017})}\BibitemShut {NoStop}%
\bibitem [{\citenamefont {Wan}\ \emph {et~al.}(2022)\citenamefont {Wan}, \citenamefont {Seemann}, \citenamefont {Tata}, \citenamefont {Andriyash}, \citenamefont {Smartsev}, \citenamefont {Kroupp},\ and\ \citenamefont {Malka}}]{Wan2022}%
  \BibitemOpen
  \bibfield  {author} {\bibinfo {author} {\bibfnamefont {Y.}~\bibnamefont {Wan}}, \bibinfo {author} {\bibfnamefont {O.}~\bibnamefont {Seemann}}, \bibinfo {author} {\bibfnamefont {S.}~\bibnamefont {Tata}}, \bibinfo {author} {\bibfnamefont {I.~A.}\ \bibnamefont {Andriyash}}, \bibinfo {author} {\bibfnamefont {S.}~\bibnamefont {Smartsev}}, \bibinfo {author} {\bibfnamefont {E.}~\bibnamefont {Kroupp}},\ and\ \bibinfo {author} {\bibfnamefont {V.}~\bibnamefont {Malka}},\ }\bibfield  {title} {\bibinfo {title} {Direct observation of relativistic broken plasma waves},\ }\href {https://doi.org/10.1038/s41567-022-01717-6} {\bibfield  {journal} {\bibinfo  {journal} {Nature Physics}\ }\textbf {\bibinfo {volume} {18}},\ \bibinfo {pages} {1186} (\bibinfo {year} {2022})}\BibitemShut {NoStop}%
\bibitem [{\citenamefont {Bruhaug}\ \emph {et~al.}(2023)\citenamefont {Bruhaug}, \citenamefont {Freeman}, \citenamefont {Rinderknecht}, \citenamefont {Neukirch}, \citenamefont {Wilde}, \citenamefont {Merrill}, \citenamefont {Rygg}, \citenamefont {Wei}, \citenamefont {Collins},\ and\ \citenamefont {Shaw}}]{Bruhaug2023}%
  \BibitemOpen
  \bibfield  {author} {\bibinfo {author} {\bibfnamefont {G.}~\bibnamefont {Bruhaug}}, \bibinfo {author} {\bibfnamefont {M.~S.}\ \bibnamefont {Freeman}}, \bibinfo {author} {\bibfnamefont {H.~G.}\ \bibnamefont {Rinderknecht}}, \bibinfo {author} {\bibfnamefont {L.~P.}\ \bibnamefont {Neukirch}}, \bibinfo {author} {\bibfnamefont {C.~H.}\ \bibnamefont {Wilde}}, \bibinfo {author} {\bibfnamefont {F.~E.}\ \bibnamefont {Merrill}}, \bibinfo {author} {\bibfnamefont {J.~R.}\ \bibnamefont {Rygg}}, \bibinfo {author} {\bibfnamefont {M.~S.}\ \bibnamefont {Wei}}, \bibinfo {author} {\bibfnamefont {G.~W.}\ \bibnamefont {Collins}},\ and\ \bibinfo {author} {\bibfnamefont {J.~L.}\ \bibnamefont {Shaw}},\ }\bibfield  {title} {\bibinfo {title} {Single-shot electron radiography using a laser–plasma accelerator},\ }\href {https://doi.org/10.1038/s41598-023-29217-4} {\bibfield  {journal} {\bibinfo  {journal} {Scientific Reports}\ }\textbf {\bibinfo {volume} {13}},\ \bibinfo {pages} {2227} (\bibinfo {year} {2023})}\BibitemShut {NoStop}%
\bibitem [{\citenamefont {Pomerantz}\ \emph {et~al.}(2014)\citenamefont {Pomerantz}, \citenamefont {McCary}, \citenamefont {Meadows}, \citenamefont {Arefiev}, \citenamefont {Bernstein}, \citenamefont {Chester}, \citenamefont {Cortez}, \citenamefont {Donovan}, \citenamefont {Dyer}, \citenamefont {Gaul}, \citenamefont {Hamilton}, \citenamefont {Kuk}, \citenamefont {Lestrade}, \citenamefont {Wang}, \citenamefont {Ditmire},\ and\ \citenamefont {Hegelich}}]{Pomerantz2014a}%
  \BibitemOpen
  \bibfield  {author} {\bibinfo {author} {\bibfnamefont {I.}~\bibnamefont {Pomerantz}}, \bibinfo {author} {\bibfnamefont {E.}~\bibnamefont {McCary}}, \bibinfo {author} {\bibfnamefont {A.~R.}\ \bibnamefont {Meadows}}, \bibinfo {author} {\bibfnamefont {A.}~\bibnamefont {Arefiev}}, \bibinfo {author} {\bibfnamefont {A.~C.}\ \bibnamefont {Bernstein}}, \bibinfo {author} {\bibfnamefont {C.}~\bibnamefont {Chester}}, \bibinfo {author} {\bibfnamefont {J.}~\bibnamefont {Cortez}}, \bibinfo {author} {\bibfnamefont {M.~E.}\ \bibnamefont {Donovan}}, \bibinfo {author} {\bibfnamefont {G.}~\bibnamefont {Dyer}}, \bibinfo {author} {\bibfnamefont {E.~W.}\ \bibnamefont {Gaul}}, \bibinfo {author} {\bibfnamefont {D.}~\bibnamefont {Hamilton}}, \bibinfo {author} {\bibfnamefont {D.}~\bibnamefont {Kuk}}, \bibinfo {author} {\bibfnamefont {A.~C.}\ \bibnamefont {Lestrade}}, \bibinfo {author} {\bibfnamefont {C.}~\bibnamefont {Wang}}, \bibinfo {author} {\bibfnamefont {T.}~\bibnamefont {Ditmire}},\ and\ \bibinfo {author} {\bibfnamefont
  {B.~M.}\ \bibnamefont {Hegelich}},\ }\bibfield  {title} {\bibinfo {title} {Ultrashort pulsed neutron source},\ }\href {https://doi.org/10.1103/PhysRevLett.113.184801} {\bibfield  {journal} {\bibinfo  {journal} {Physical Review Letters}\ }\textbf {\bibinfo {volume} {113}},\ \bibinfo {pages} {184801} (\bibinfo {year} {2014})}\BibitemShut {NoStop}%
\bibitem [{\citenamefont {G{\"u}nther}\ \emph {et~al.}(2022)\citenamefont {G{\"u}nther}, \citenamefont {Rosmej}, \citenamefont {Tavana}, \citenamefont {Gyrdymov}, \citenamefont {Skobliakov}, \citenamefont {Kantsyrev}, \citenamefont {Z{\"a}hter}, \citenamefont {Borisenko}, \citenamefont {Pukhov},\ and\ \citenamefont {Andreev}}]{Gunther2022}%
  \BibitemOpen
  \bibfield  {author} {\bibinfo {author} {\bibfnamefont {M.~M.}\ \bibnamefont {G{\"u}nther}}, \bibinfo {author} {\bibfnamefont {O.~N.}\ \bibnamefont {Rosmej}}, \bibinfo {author} {\bibfnamefont {P.}~\bibnamefont {Tavana}}, \bibinfo {author} {\bibfnamefont {M.}~\bibnamefont {Gyrdymov}}, \bibinfo {author} {\bibfnamefont {A.}~\bibnamefont {Skobliakov}}, \bibinfo {author} {\bibfnamefont {A.}~\bibnamefont {Kantsyrev}}, \bibinfo {author} {\bibfnamefont {S.}~\bibnamefont {Z{\"a}hter}}, \bibinfo {author} {\bibfnamefont {N.~G.}\ \bibnamefont {Borisenko}}, \bibinfo {author} {\bibfnamefont {A.}~\bibnamefont {Pukhov}},\ and\ \bibinfo {author} {\bibfnamefont {N.~E.}\ \bibnamefont {Andreev}},\ }\bibfield  {title} {\bibinfo {title} {Forward-looking insights in laser-generated ultra-intense $\gamma$-ray and neutron sources for nuclear application and science},\ }\href {https://doi.org/10.1038/s41467-021-27694-7} {\bibfield  {journal} {\bibinfo  {journal} {Nature Communications}\ }\textbf {\bibinfo {volume} {13}},\ \bibinfo
  {pages} {170} (\bibinfo {year} {2022})}\BibitemShut {NoStop}%
\bibitem [{\citenamefont {Cohen}\ \emph {et~al.}(2024{\natexlab{a}})\citenamefont {Cohen}, \citenamefont {Cohen}, \citenamefont {Levinson}, \citenamefont {Elkind}, \citenamefont {Rakovsky}, \citenamefont {Levanon}, \citenamefont {Michaeli}, \citenamefont {Cohen}, \citenamefont {Beck},\ and\ \citenamefont {Pomerantz}}]{Cohen2024_neutrons}%
  \BibitemOpen
  \bibfield  {author} {\bibinfo {author} {\bibfnamefont {I.}~\bibnamefont {Cohen}}, \bibinfo {author} {\bibfnamefont {T.}~\bibnamefont {Cohen}}, \bibinfo {author} {\bibfnamefont {A.}~\bibnamefont {Levinson}}, \bibinfo {author} {\bibfnamefont {M.}~\bibnamefont {Elkind}}, \bibinfo {author} {\bibfnamefont {Y.}~\bibnamefont {Rakovsky}}, \bibinfo {author} {\bibfnamefont {A.}~\bibnamefont {Levanon}}, \bibinfo {author} {\bibfnamefont {D.}~\bibnamefont {Michaeli}}, \bibinfo {author} {\bibfnamefont {E.}~\bibnamefont {Cohen}}, \bibinfo {author} {\bibfnamefont {A.}~\bibnamefont {Beck}},\ and\ \bibinfo {author} {\bibfnamefont {I.}~\bibnamefont {Pomerantz}},\ }\bibfield  {title} {\bibinfo {title} {Accumulated laser-photoneutron generation},\ }\bibfield  {journal} {\bibinfo  {journal} {European Physical Journal Plus}\ }\textbf {\bibinfo {volume} {139}},\ \href {https://doi.org/10.1140/epjp/s13360-024-05387-6} {10.1140/epjp/s13360-024-05387-6} (\bibinfo {year} {2024}{\natexlab{a}})\BibitemShut {NoStop}%
\bibitem [{\citenamefont {Kneip}\ \emph {et~al.}(2008)\citenamefont {Kneip}, \citenamefont {Nagel}, \citenamefont {Bellei}, \citenamefont {Bourgeois}, \citenamefont {Dangor}, \citenamefont {Gopal}, \citenamefont {Heathcote}, \citenamefont {Mangles}, \citenamefont {Marqu\`es}, \citenamefont {Maksimchuk}, \citenamefont {Nilson}, \citenamefont {Phuoc}, \citenamefont {Reed}, \citenamefont {Tzoufras}, \citenamefont {Tsung}, \citenamefont {Willingale}, \citenamefont {Mori}, \citenamefont {Rousse}, \citenamefont {Krushelnick},\ and\ \citenamefont {Najmudin}}]{Kneip2008}%
  \BibitemOpen
  \bibfield  {author} {\bibinfo {author} {\bibfnamefont {S.}~\bibnamefont {Kneip}}, \bibinfo {author} {\bibfnamefont {S.~R.}\ \bibnamefont {Nagel}}, \bibinfo {author} {\bibfnamefont {C.}~\bibnamefont {Bellei}}, \bibinfo {author} {\bibfnamefont {N.}~\bibnamefont {Bourgeois}}, \bibinfo {author} {\bibfnamefont {A.~E.}\ \bibnamefont {Dangor}}, \bibinfo {author} {\bibfnamefont {A.}~\bibnamefont {Gopal}}, \bibinfo {author} {\bibfnamefont {R.}~\bibnamefont {Heathcote}}, \bibinfo {author} {\bibfnamefont {S.~P.~D.}\ \bibnamefont {Mangles}}, \bibinfo {author} {\bibfnamefont {J.~R.}\ \bibnamefont {Marqu\`es}}, \bibinfo {author} {\bibfnamefont {A.}~\bibnamefont {Maksimchuk}}, \bibinfo {author} {\bibfnamefont {P.~M.}\ \bibnamefont {Nilson}}, \bibinfo {author} {\bibfnamefont {K.~T.}\ \bibnamefont {Phuoc}}, \bibinfo {author} {\bibfnamefont {S.}~\bibnamefont {Reed}}, \bibinfo {author} {\bibfnamefont {M.}~\bibnamefont {Tzoufras}}, \bibinfo {author} {\bibfnamefont {F.~S.}\ \bibnamefont {Tsung}}, \bibinfo {author}
  {\bibfnamefont {L.}~\bibnamefont {Willingale}}, \bibinfo {author} {\bibfnamefont {W.~B.}\ \bibnamefont {Mori}}, \bibinfo {author} {\bibfnamefont {A.}~\bibnamefont {Rousse}}, \bibinfo {author} {\bibfnamefont {K.}~\bibnamefont {Krushelnick}},\ and\ \bibinfo {author} {\bibfnamefont {Z.}~\bibnamefont {Najmudin}},\ }\bibfield  {title} {\bibinfo {title} {Observation of synchrotron radiation from electrons accelerated in a petawatt-laser-generated plasma cavity},\ }\href {https://doi.org/10.1103/PhysRevLett.100.105006} {\bibfield  {journal} {\bibinfo  {journal} {Phys. Rev. Lett.}\ }\textbf {\bibinfo {volume} {100}},\ \bibinfo {pages} {105006} (\bibinfo {year} {2008})}\BibitemShut {NoStop}%
\bibitem [{\citenamefont {Cikhardt}\ \emph {et~al.}(2024)\citenamefont {Cikhardt}, \citenamefont {Gyrdymov}, \citenamefont {Zähter}, \citenamefont {Tavana}, \citenamefont {Günther}, \citenamefont {Bukharskii}, \citenamefont {Borisenko}, \citenamefont {Jacoby}, \citenamefont {Shen}, \citenamefont {Pukhov}, \citenamefont {Andreev},\ and\ \citenamefont {Rosmej}}]{Cikhardt2024}%
  \BibitemOpen
  \bibfield  {author} {\bibinfo {author} {\bibfnamefont {J.}~\bibnamefont {Cikhardt}}, \bibinfo {author} {\bibfnamefont {M.}~\bibnamefont {Gyrdymov}}, \bibinfo {author} {\bibfnamefont {S.}~\bibnamefont {Zähter}}, \bibinfo {author} {\bibfnamefont {P.}~\bibnamefont {Tavana}}, \bibinfo {author} {\bibfnamefont {M.~M.}\ \bibnamefont {Günther}}, \bibinfo {author} {\bibfnamefont {N.}~\bibnamefont {Bukharskii}}, \bibinfo {author} {\bibfnamefont {N.}~\bibnamefont {Borisenko}}, \bibinfo {author} {\bibfnamefont {J.}~\bibnamefont {Jacoby}}, \bibinfo {author} {\bibfnamefont {X.~F.}\ \bibnamefont {Shen}}, \bibinfo {author} {\bibfnamefont {A.}~\bibnamefont {Pukhov}}, \bibinfo {author} {\bibfnamefont {N.~E.}\ \bibnamefont {Andreev}},\ and\ \bibinfo {author} {\bibfnamefont {O.~N.}\ \bibnamefont {Rosmej}},\ }\bibfield  {title} {\bibinfo {title} {Characterization of bright betatron radiation generated by direct laser acceleration of electrons in plasma of near critical density},\ }\href {https://doi.org/10.1063/5.0181119}
  {\bibfield  {journal} {\bibinfo  {journal} {Matter and Radiation at Extremes}\ }\textbf {\bibinfo {volume} {9}},\ \bibinfo {pages} {027201} (\bibinfo {year} {2024})}\BibitemShut {NoStop}%
\bibitem [{\citenamefont {Meir}\ \emph {et~al.}(2024)\citenamefont {Meir}, \citenamefont {Cohen}, \citenamefont {Tangtartharakul}, \citenamefont {Cohen}, \citenamefont {Fraenkel}, \citenamefont {Arefiev},\ and\ \citenamefont {Pomerantz}}]{Meir2024}%
  \BibitemOpen
  \bibfield  {author} {\bibinfo {author} {\bibfnamefont {T.}~\bibnamefont {Meir}}, \bibinfo {author} {\bibfnamefont {I.}~\bibnamefont {Cohen}}, \bibinfo {author} {\bibfnamefont {K.}~\bibnamefont {Tangtartharakul}}, \bibinfo {author} {\bibfnamefont {T.}~\bibnamefont {Cohen}}, \bibinfo {author} {\bibfnamefont {M.}~\bibnamefont {Fraenkel}}, \bibinfo {author} {\bibfnamefont {A.~V.}\ \bibnamefont {Arefiev}},\ and\ \bibinfo {author} {\bibfnamefont {I.}~\bibnamefont {Pomerantz}},\ }\bibfield  {title} {\bibinfo {title} {Plasma-guided compton source},\ }\href {https://doi.org/10.1103/PhysRevApplied.22.044004} {\bibfield  {journal} {\bibinfo  {journal} {Phys. Rev. Appl.}\ }\textbf {\bibinfo {volume} {22}},\ \bibinfo {pages} {044004} (\bibinfo {year} {2024})}\BibitemShut {NoStop}%
\bibitem [{\citenamefont {Babjak}\ and\ \citenamefont {Vranic}(2025)}]{Babjak2025}%
  \BibitemOpen
  \bibfield  {author} {\bibinfo {author} {\bibfnamefont {R.}~\bibnamefont {Babjak}}\ and\ \bibinfo {author} {\bibfnamefont {M.}~\bibnamefont {Vranic}},\ }\bibfield  {title} {\bibinfo {title} {Betatron radiation emitted during the direct laser acceleration of electrons in underdense plasmas},\ }\href {https://doi.org/10.1088/1361-6587/adf50b} {\bibfield  {journal} {\bibinfo  {journal} {Plasma Physics and Controlled Fusion}\ }\textbf {\bibinfo {volume} {67}},\ \bibinfo {pages} {085019} (\bibinfo {year} {2025})}\BibitemShut {NoStop}%
\bibitem [{\citenamefont {Tangtartharakul}\ \emph {et~al.}(2025)\citenamefont {Tangtartharakul}, \citenamefont {Fauvel}, \citenamefont {Meir}, \citenamefont {Condamine}, \citenamefont {Weber}, \citenamefont {Pomerantz}, \citenamefont {Manuel},\ and\ \citenamefont {Arefiev}}]{Tangtartharakul_2025}%
  \BibitemOpen
  \bibfield  {author} {\bibinfo {author} {\bibfnamefont {K.}~\bibnamefont {Tangtartharakul}}, \bibinfo {author} {\bibfnamefont {G.}~\bibnamefont {Fauvel}}, \bibinfo {author} {\bibfnamefont {T.}~\bibnamefont {Meir}}, \bibinfo {author} {\bibfnamefont {F.~P.}\ \bibnamefont {Condamine}}, \bibinfo {author} {\bibfnamefont {S.}~\bibnamefont {Weber}}, \bibinfo {author} {\bibfnamefont {I.}~\bibnamefont {Pomerantz}}, \bibinfo {author} {\bibfnamefont {M.}~\bibnamefont {Manuel}},\ and\ \bibinfo {author} {\bibfnamefont {A.}~\bibnamefont {Arefiev}},\ }\bibfield  {title} {\bibinfo {title} {Collimated gamma-ray emission enabled by efficient direct laser acceleration},\ }\href {https://doi.org/10.1088/1367-2630/adb3c1} {\bibfield  {journal} {\bibinfo  {journal} {New Journal of Physics}\ }\textbf {\bibinfo {volume} {27}},\ \bibinfo {pages} {023024} (\bibinfo {year} {2025})}\BibitemShut {NoStop}%
\bibitem [{\citenamefont {Meir}\ \emph {et~al.}(2025)\citenamefont {Meir}, \citenamefont {Cohen}, \citenamefont {Tangtartharakul}, \citenamefont {Cohen}, \citenamefont {Fraenkel}, \citenamefont {Arefiev},\ and\ \citenamefont {Pomerantz}}]{Proceeding2025}%
  \BibitemOpen
  \bibfield  {author} {\bibinfo {author} {\bibfnamefont {T.}~\bibnamefont {Meir}}, \bibinfo {author} {\bibfnamefont {I.}~\bibnamefont {Cohen}}, \bibinfo {author} {\bibfnamefont {K.}~\bibnamefont {Tangtartharakul}}, \bibinfo {author} {\bibfnamefont {T.}~\bibnamefont {Cohen}}, \bibinfo {author} {\bibfnamefont {M.}~\bibnamefont {Fraenkel}}, \bibinfo {author} {\bibfnamefont {A.~V.}\ \bibnamefont {Arefiev}},\ and\ \bibinfo {author} {\bibfnamefont {I.}~\bibnamefont {Pomerantz}},\ }\bibfield  {title} {\bibinfo {title} {{Inverse Compton scattering with direct laser acceleration}},\ }in\ \href {https://doi.org/10.1117/12.3056200} {\emph {\bibinfo {booktitle} {Compact Radiation Sources from EUV to Gamma-rays: Development and Applications II}}},\ Vol.\ \bibinfo {volume} {13537},\ \bibinfo {editor} {edited by\ \bibinfo {editor} {\bibfnamefont {C.~S.}\ \bibnamefont {Menoni}}\ and\ \bibinfo {editor} {\bibfnamefont {J.}~\bibnamefont {Nejdl}}},\ \bibinfo {organization} {International Society for Optics and Photonics}\
  (\bibinfo  {publisher} {SPIE},\ \bibinfo {year} {2025})\ p.\ \bibinfo {pages} {1353708}\BibitemShut {NoStop}%
\bibitem [{\citenamefont {Cole}\ \emph {et~al.}(2018)\citenamefont {Cole}, \citenamefont {Behm}, \citenamefont {Gerstmayr}, \citenamefont {Blackburn}, \citenamefont {Wood}, \citenamefont {Baird}, \citenamefont {Duff}, \citenamefont {Harvey}, \citenamefont {Ilderton}, \citenamefont {Joglekar}, \citenamefont {Krushelnick}, \citenamefont {Kuschel}, \citenamefont {Marklund}, \citenamefont {McKenna}, \citenamefont {Murphy}, \citenamefont {Poder}, \citenamefont {Ridgers}, \citenamefont {Samarin}, \citenamefont {Sarri}, \citenamefont {Symes}, \citenamefont {Thomas}, \citenamefont {Warwick}, \citenamefont {Zepf}, \citenamefont {Najmudin},\ and\ \citenamefont {Mangles}}]{Cole2018}%
  \BibitemOpen
  \bibfield  {author} {\bibinfo {author} {\bibfnamefont {J.~M.}\ \bibnamefont {Cole}}, \bibinfo {author} {\bibfnamefont {K.~T.}\ \bibnamefont {Behm}}, \bibinfo {author} {\bibfnamefont {E.}~\bibnamefont {Gerstmayr}}, \bibinfo {author} {\bibfnamefont {T.~G.}\ \bibnamefont {Blackburn}}, \bibinfo {author} {\bibfnamefont {J.~C.}\ \bibnamefont {Wood}}, \bibinfo {author} {\bibfnamefont {C.~D.}\ \bibnamefont {Baird}}, \bibinfo {author} {\bibfnamefont {M.~J.}\ \bibnamefont {Duff}}, \bibinfo {author} {\bibfnamefont {C.}~\bibnamefont {Harvey}}, \bibinfo {author} {\bibfnamefont {A.}~\bibnamefont {Ilderton}}, \bibinfo {author} {\bibfnamefont {A.~S.}\ \bibnamefont {Joglekar}}, \bibinfo {author} {\bibfnamefont {K.}~\bibnamefont {Krushelnick}}, \bibinfo {author} {\bibfnamefont {S.}~\bibnamefont {Kuschel}}, \bibinfo {author} {\bibfnamefont {M.}~\bibnamefont {Marklund}}, \bibinfo {author} {\bibfnamefont {P.}~\bibnamefont {McKenna}}, \bibinfo {author} {\bibfnamefont {C.~D.}\ \bibnamefont {Murphy}}, \bibinfo {author}
  {\bibfnamefont {K.}~\bibnamefont {Poder}}, \bibinfo {author} {\bibfnamefont {C.~P.}\ \bibnamefont {Ridgers}}, \bibinfo {author} {\bibfnamefont {G.~M.}\ \bibnamefont {Samarin}}, \bibinfo {author} {\bibfnamefont {G.}~\bibnamefont {Sarri}}, \bibinfo {author} {\bibfnamefont {D.~R.}\ \bibnamefont {Symes}}, \bibinfo {author} {\bibfnamefont {A.~G.~R.}\ \bibnamefont {Thomas}}, \bibinfo {author} {\bibfnamefont {J.}~\bibnamefont {Warwick}}, \bibinfo {author} {\bibfnamefont {M.}~\bibnamefont {Zepf}}, \bibinfo {author} {\bibfnamefont {Z.}~\bibnamefont {Najmudin}},\ and\ \bibinfo {author} {\bibfnamefont {S.~P.~D.}\ \bibnamefont {Mangles}},\ }\bibfield  {title} {\bibinfo {title} {Experimental evidence of radiation reaction in the collision of a high-intensity laser pulse with a laser-wakefield accelerated electron beam},\ }\href {https://doi.org/10.1103/PhysRevX.8.011020} {\bibfield  {journal} {\bibinfo  {journal} {Phys. Rev. X}\ }\textbf {\bibinfo {volume} {8}},\ \bibinfo {pages} {011020} (\bibinfo {year}
  {2018})}\BibitemShut {NoStop}%
\bibitem [{\citenamefont {Blackburn}(2020)}]{blackburn2020radiation}%
  \BibitemOpen
  \bibfield  {author} {\bibinfo {author} {\bibfnamefont {T.}~\bibnamefont {Blackburn}},\ }\bibfield  {title} {\bibinfo {title} {Radiation reaction in electron--beam interactions with high-intensity lasers},\ }\href@noop {} {\bibfield  {journal} {\bibinfo  {journal} {Reviews of Modern Plasma Physics}\ }\textbf {\bibinfo {volume} {4}},\ \bibinfo {pages} {5} (\bibinfo {year} {2020})}\BibitemShut {NoStop}%
\bibitem [{\citenamefont {Gonoskov}\ \emph {et~al.}(2021)\citenamefont {Gonoskov}, \citenamefont {Blackburn}, \citenamefont {Marklund},\ and\ \citenamefont {Bulanov}}]{Gonoskov2021Charged}%
  \BibitemOpen
  \bibfield  {author} {\bibinfo {author} {\bibfnamefont {A.}~\bibnamefont {Gonoskov}}, \bibinfo {author} {\bibfnamefont {T.}~\bibnamefont {Blackburn}}, \bibinfo {author} {\bibfnamefont {M.}~\bibnamefont {Marklund}},\ and\ \bibinfo {author} {\bibfnamefont {S.}~\bibnamefont {Bulanov}},\ }\bibfield  {title} {\bibinfo {title} {Charged particle motion and radiation in strong electromagnetic fields},\ }\bibfield  {journal} {\bibinfo  {journal} {Reviews of Modern Physics}\ }\href {https://doi.org/10.1103/RevModPhys.94.045001} {10.1103/RevModPhys.94.045001} (\bibinfo {year} {2021})\BibitemShut {NoStop}%
\bibitem [{\citenamefont {Arefiev}\ \emph {et~al.}(2015)\citenamefont {Arefiev}, \citenamefont {Robinson},\ and\ \citenamefont {Khudik}}]{Arefiev2015}%
  \BibitemOpen
  \bibfield  {author} {\bibinfo {author} {\bibfnamefont {A.~V.}\ \bibnamefont {Arefiev}}, \bibinfo {author} {\bibfnamefont {A.~P.}\ \bibnamefont {Robinson}},\ and\ \bibinfo {author} {\bibfnamefont {V.~N.}\ \bibnamefont {Khudik}},\ }\bibfield  {title} {\bibinfo {title} {Novel aspects of direct laser acceleration of relativistic electrons},\ }\bibfield  {journal} {\bibinfo  {journal} {Journal of Plasma Physics}\ }\textbf {\bibinfo {volume} {81}},\ \href {https://doi.org/10.1017/S0022377815000434} {10.1017/S0022377815000434} (\bibinfo {year} {2015})\BibitemShut {NoStop}%
\bibitem [{\citenamefont {Malka}\ \emph {et~al.}(1997)\citenamefont {Malka}, \citenamefont {Lefebvre},\ and\ \citenamefont {Miquel}}]{Malka1997}%
  \BibitemOpen
  \bibfield  {author} {\bibinfo {author} {\bibfnamefont {G.}~\bibnamefont {Malka}}, \bibinfo {author} {\bibfnamefont {E.}~\bibnamefont {Lefebvre}},\ and\ \bibinfo {author} {\bibfnamefont {J.~L.}\ \bibnamefont {Miquel}},\ }\bibfield  {title} {\bibinfo {title} {Experimental observation of electrons accelerated in vacuum to relativistic energies by a high-intensity laser},\ }\href {https://doi.org/10.1103/PhysRevLett.78.3314} {\bibfield  {journal} {\bibinfo  {journal} {Physical Review Letters}\ }\textbf {\bibinfo {volume} {78}},\ \bibinfo {pages} {3314} (\bibinfo {year} {1997})}\BibitemShut {NoStop}%
\bibitem [{\citenamefont {Gahn}\ \emph {et~al.}(1999)\citenamefont {Gahn}, \citenamefont {Tsakiris}, \citenamefont {Pukhov}, \citenamefont {Meyer-Ter-Vehn}, \citenamefont {Pretzler}, \citenamefont {Thirolf}, \citenamefont {Habs},\ and\ \citenamefont {Witte}}]{Gahn1999}%
  \BibitemOpen
  \bibfield  {author} {\bibinfo {author} {\bibfnamefont {C.}~\bibnamefont {Gahn}}, \bibinfo {author} {\bibfnamefont {G.~D.}\ \bibnamefont {Tsakiris}}, \bibinfo {author} {\bibfnamefont {A.}~\bibnamefont {Pukhov}}, \bibinfo {author} {\bibfnamefont {J.}~\bibnamefont {Meyer-Ter-Vehn}}, \bibinfo {author} {\bibfnamefont {G.}~\bibnamefont {Pretzler}}, \bibinfo {author} {\bibfnamefont {P.}~\bibnamefont {Thirolf}}, \bibinfo {author} {\bibfnamefont {D.}~\bibnamefont {Habs}},\ and\ \bibinfo {author} {\bibfnamefont {K.~J.}\ \bibnamefont {Witte}},\ }\bibfield  {title} {\bibinfo {title} {{Multi-MeV electron beam generation by direct laser acceleration in high-density plasma channels}},\ }\href {https://doi.org/10.1103/PhysRevLett.83.4772} {\bibfield  {journal} {\bibinfo  {journal} {Physical Review Letters}\ }\textbf {\bibinfo {volume} {83}},\ \bibinfo {pages} {4772} (\bibinfo {year} {1999})}\BibitemShut {NoStop}%
\bibitem [{\citenamefont {Shaw}\ \emph {et~al.}(2018)\citenamefont {Shaw}, \citenamefont {Lemos}, \citenamefont {Marsh}, \citenamefont {Froula},\ and\ \citenamefont {Joshi}}]{Shaw_2018}%
  \BibitemOpen
  \bibfield  {author} {\bibinfo {author} {\bibfnamefont {J.~L.}\ \bibnamefont {Shaw}}, \bibinfo {author} {\bibfnamefont {N.}~\bibnamefont {Lemos}}, \bibinfo {author} {\bibfnamefont {K.~A.}\ \bibnamefont {Marsh}}, \bibinfo {author} {\bibfnamefont {D.~H.}\ \bibnamefont {Froula}},\ and\ \bibinfo {author} {\bibfnamefont {C.}~\bibnamefont {Joshi}},\ }\bibfield  {title} {\bibinfo {title} {Experimental signatures of direct-laser-acceleration-assisted laser wakefield acceleration},\ }\href {https://doi.org/10.1088/1361-6587/aaade1} {\bibfield  {journal} {\bibinfo  {journal} {Plasma Physics and Controlled Fusion}\ }\textbf {\bibinfo {volume} {60}},\ \bibinfo {pages} {044012} (\bibinfo {year} {2018})}\BibitemShut {NoStop}%
\bibitem [{\citenamefont {Hussein}\ \emph {et~al.}(2021)\citenamefont {Hussein}, \citenamefont {Arefiev}, \citenamefont {Batson}, \citenamefont {Chen}, \citenamefont {Craxton}, \citenamefont {Davies}, \citenamefont {Froula}, \citenamefont {Gong}, \citenamefont {Haberberger}, \citenamefont {Ma}, \citenamefont {Nilson}, \citenamefont {Theobald}, \citenamefont {Wang}, \citenamefont {Weichman}, \citenamefont {Williams},\ and\ \citenamefont {Willingale}}]{Hussein_2021}%
  \BibitemOpen
  \bibfield  {author} {\bibinfo {author} {\bibfnamefont {A.~E.}\ \bibnamefont {Hussein}}, \bibinfo {author} {\bibfnamefont {A.~V.}\ \bibnamefont {Arefiev}}, \bibinfo {author} {\bibfnamefont {T.}~\bibnamefont {Batson}}, \bibinfo {author} {\bibfnamefont {H.}~\bibnamefont {Chen}}, \bibinfo {author} {\bibfnamefont {R.~S.}\ \bibnamefont {Craxton}}, \bibinfo {author} {\bibfnamefont {A.~S.}\ \bibnamefont {Davies}}, \bibinfo {author} {\bibfnamefont {D.~H.}\ \bibnamefont {Froula}}, \bibinfo {author} {\bibfnamefont {Z.}~\bibnamefont {Gong}}, \bibinfo {author} {\bibfnamefont {D.}~\bibnamefont {Haberberger}}, \bibinfo {author} {\bibfnamefont {Y.}~\bibnamefont {Ma}}, \bibinfo {author} {\bibfnamefont {P.~M.}\ \bibnamefont {Nilson}}, \bibinfo {author} {\bibfnamefont {W.}~\bibnamefont {Theobald}}, \bibinfo {author} {\bibfnamefont {T.}~\bibnamefont {Wang}}, \bibinfo {author} {\bibfnamefont {K.}~\bibnamefont {Weichman}}, \bibinfo {author} {\bibfnamefont {G.~J.}\ \bibnamefont {Williams}},\ and\ \bibinfo {author} {\bibfnamefont
  {L.}~\bibnamefont {Willingale}},\ }\bibfield  {title} {\bibinfo {title} {Towards the optimisation of direct laser acceleration},\ }\href {https://doi.org/10.1088/1367-2630/abdf9a} {\bibfield  {journal} {\bibinfo  {journal} {New Journal of Physics}\ }\textbf {\bibinfo {volume} {23}},\ \bibinfo {pages} {023031} (\bibinfo {year} {2021})}\BibitemShut {NoStop}%
\bibitem [{\citenamefont {Cohen}\ \emph {et~al.}(2024{\natexlab{b}})\citenamefont {Cohen}, \citenamefont {Meir}, \citenamefont {Tangtartharakul}, \citenamefont {Perelmutter}, \citenamefont {Elkind}, \citenamefont {Gershuni}, \citenamefont {Levanon}, \citenamefont {Arefiev},\ and\ \citenamefont {Pomerantz}}]{Cohen2024}%
  \BibitemOpen
  \bibfield  {author} {\bibinfo {author} {\bibfnamefont {I.}~\bibnamefont {Cohen}}, \bibinfo {author} {\bibfnamefont {T.}~\bibnamefont {Meir}}, \bibinfo {author} {\bibfnamefont {K.}~\bibnamefont {Tangtartharakul}}, \bibinfo {author} {\bibfnamefont {L.}~\bibnamefont {Perelmutter}}, \bibinfo {author} {\bibfnamefont {M.}~\bibnamefont {Elkind}}, \bibinfo {author} {\bibfnamefont {Y.}~\bibnamefont {Gershuni}}, \bibinfo {author} {\bibfnamefont {A.}~\bibnamefont {Levanon}}, \bibinfo {author} {\bibfnamefont {A.~V.}\ \bibnamefont {Arefiev}},\ and\ \bibinfo {author} {\bibfnamefont {I.}~\bibnamefont {Pomerantz}},\ }\bibfield  {title} {\bibinfo {title} {Undepleted direct laser acceleration},\ }\href {https://doi.org/10.1126/sciadv.adk1947} {\bibfield  {journal} {\bibinfo  {journal} {Science Advances}\ }\textbf {\bibinfo {volume} {10}},\ \bibinfo {pages} {eadk1947} (\bibinfo {year} {2024}{\natexlab{b}})},\ \Eprint {https://arxiv.org/abs/https://www.science.org/doi/pdf/10.1126/sciadv.adk1947}
  {https://www.science.org/doi/pdf/10.1126/sciadv.adk1947} \BibitemShut {NoStop}%
\bibitem [{\citenamefont {Babjak}\ \emph {et~al.}(2024)\citenamefont {Babjak}, \citenamefont {Willingale}, \citenamefont {Arefiev},\ and\ \citenamefont {Vranic}}]{Babjak2024PRL}%
  \BibitemOpen
  \bibfield  {author} {\bibinfo {author} {\bibfnamefont {R.}~\bibnamefont {Babjak}}, \bibinfo {author} {\bibfnamefont {L.}~\bibnamefont {Willingale}}, \bibinfo {author} {\bibfnamefont {A.}~\bibnamefont {Arefiev}},\ and\ \bibinfo {author} {\bibfnamefont {M.}~\bibnamefont {Vranic}},\ }\bibfield  {title} {\bibinfo {title} {Direct laser acceleration in underdense plasmas with multi-pw lasers: A path to high-charge, gev-class electron bunches},\ }\href {https://doi.org/10.1103/PhysRevLett.132.125001} {\bibfield  {journal} {\bibinfo  {journal} {Phys. Rev. Lett.}\ }\textbf {\bibinfo {volume} {132}},\ \bibinfo {pages} {125001} (\bibinfo {year} {2024})}\BibitemShut {NoStop}%
\bibitem [{\citenamefont {Rosmej}\ \emph {et~al.}(2025)\citenamefont {Rosmej}, \citenamefont {Gyrdymov}, \citenamefont {Andreev}, \citenamefont {Tavana}, \citenamefont {Popov}, \citenamefont {Borisenko}, \citenamefont {Gromov}, \citenamefont {Gus'kov}, \citenamefont {Yakhin}, \citenamefont {Vegunova} \emph {et~al.}}]{Rosmej2025}%
  \BibitemOpen
  \bibfield  {author} {\bibinfo {author} {\bibfnamefont {O.~N.}\ \bibnamefont {Rosmej}}, \bibinfo {author} {\bibfnamefont {M.}~\bibnamefont {Gyrdymov}}, \bibinfo {author} {\bibfnamefont {N.~E.}\ \bibnamefont {Andreev}}, \bibinfo {author} {\bibfnamefont {P.}~\bibnamefont {Tavana}}, \bibinfo {author} {\bibfnamefont {V.}~\bibnamefont {Popov}}, \bibinfo {author} {\bibfnamefont {N.~G.}\ \bibnamefont {Borisenko}}, \bibinfo {author} {\bibfnamefont {A.~I.}\ \bibnamefont {Gromov}}, \bibinfo {author} {\bibfnamefont {S.~Y.}\ \bibnamefont {Gus'kov}}, \bibinfo {author} {\bibfnamefont {R.}~\bibnamefont {Yakhin}}, \bibinfo {author} {\bibfnamefont {G.~A.}\ \bibnamefont {Vegunova}}, \emph {et~al.},\ }\bibfield  {title} {\bibinfo {title} {Advanced plasma target from pre-ionized low-density foam for effective and robust direct laser acceleration of electrons},\ }\href {https://doi.org/10.1017/hpl.2024.85} {\bibfield  {journal} {\bibinfo  {journal} {High Power Laser Science and Engineering}\ }\textbf {\bibinfo {volume} {13}},\
  \bibinfo {pages} {e3} (\bibinfo {year} {2025})}\BibitemShut {NoStop}%
\bibitem [{\citenamefont {Pukhov}\ \emph {et~al.}(1999)\citenamefont {Pukhov}, \citenamefont {Sheng},\ and\ \citenamefont {Meyer-ter Vehn}}]{pukhov1999particle}%
  \BibitemOpen
  \bibfield  {author} {\bibinfo {author} {\bibfnamefont {A.}~\bibnamefont {Pukhov}}, \bibinfo {author} {\bibfnamefont {Z.-M.}\ \bibnamefont {Sheng}},\ and\ \bibinfo {author} {\bibfnamefont {J.}~\bibnamefont {Meyer-ter Vehn}},\ }\bibfield  {title} {\bibinfo {title} {Particle acceleration in relativistic laser channels},\ }\href@noop {} {\bibfield  {journal} {\bibinfo  {journal} {Physics of Plasmas}\ }\textbf {\bibinfo {volume} {6}},\ \bibinfo {pages} {2847} (\bibinfo {year} {1999})}\BibitemShut {NoStop}%
\bibitem [{\citenamefont {Khudik}\ \emph {et~al.}(2016)\citenamefont {Khudik}, \citenamefont {Arefiev}, \citenamefont {Zhang},\ and\ \citenamefont {Shvets}}]{Khudik2016Scaling}%
  \BibitemOpen
  \bibfield  {author} {\bibinfo {author} {\bibfnamefont {V.}~\bibnamefont {Khudik}}, \bibinfo {author} {\bibfnamefont {A.}~\bibnamefont {Arefiev}}, \bibinfo {author} {\bibfnamefont {X.}~\bibnamefont {Zhang}},\ and\ \bibinfo {author} {\bibfnamefont {G.}~\bibnamefont {Shvets}},\ }\bibfield  {title} {\bibinfo {title} {Universal scalings for laser acceleration of electrons in ion channels},\ }\href {https://doi.org/10.1063/1.4964901} {\bibfield  {journal} {\bibinfo  {journal} {Physics of Plasmas}\ }\textbf {\bibinfo {volume} {23}},\ \bibinfo {pages} {103108} (\bibinfo {year} {2016})}\BibitemShut {NoStop}%
\bibitem [{\citenamefont {Najmudin}\ \emph {et~al.}(2003)\citenamefont {Najmudin}, \citenamefont {Krushelnick}, \citenamefont {Tatarakis}, \citenamefont {Clark}, \citenamefont {Danson}, \citenamefont {Malka}, \citenamefont {Neely}, \citenamefont {Santala},\ and\ \citenamefont {Dangor}}]{Najmudin2003}%
  \BibitemOpen
  \bibfield  {author} {\bibinfo {author} {\bibfnamefont {Z.}~\bibnamefont {Najmudin}}, \bibinfo {author} {\bibfnamefont {K.}~\bibnamefont {Krushelnick}}, \bibinfo {author} {\bibfnamefont {M.}~\bibnamefont {Tatarakis}}, \bibinfo {author} {\bibfnamefont {E.~L.}\ \bibnamefont {Clark}}, \bibinfo {author} {\bibfnamefont {C.~N.}\ \bibnamefont {Danson}}, \bibinfo {author} {\bibfnamefont {V.}~\bibnamefont {Malka}}, \bibinfo {author} {\bibfnamefont {D.}~\bibnamefont {Neely}}, \bibinfo {author} {\bibfnamefont {M.~I.~K.}\ \bibnamefont {Santala}},\ and\ \bibinfo {author} {\bibfnamefont {A.~E.}\ \bibnamefont {Dangor}},\ }\bibfield  {title} {\bibinfo {title} {The effect of high intensity laser propagation instabilities on channel formation in underdense plasmas},\ }\href {https://doi.org/10.1063/1.1534585} {\bibfield  {journal} {\bibinfo  {journal} {Physics of Plasmas}\ }\textbf {\bibinfo {volume} {10}},\ \bibinfo {pages} {438} (\bibinfo {year} {2003})}\BibitemShut {NoStop}%
\bibitem [{\citenamefont {Esarey}\ \emph {et~al.}(2009)\citenamefont {Esarey}, \citenamefont {Schroeder},\ and\ \citenamefont {Leemans}}]{Esarey2009}%
  \BibitemOpen
  \bibfield  {author} {\bibinfo {author} {\bibfnamefont {E.}~\bibnamefont {Esarey}}, \bibinfo {author} {\bibfnamefont {C.~B.}\ \bibnamefont {Schroeder}},\ and\ \bibinfo {author} {\bibfnamefont {W.~P.}\ \bibnamefont {Leemans}},\ }\bibfield  {title} {\bibinfo {title} {Physics of laser-driven plasma-based electron accelerators},\ }\href {https://doi.org/10.1103/RevModPhys.81.1229} {\bibfield  {journal} {\bibinfo  {journal} {Reviews of Modern Physics}\ }\textbf {\bibinfo {volume} {81}},\ \bibinfo {pages} {1229} (\bibinfo {year} {2009})}\BibitemShut {NoStop}%
\bibitem [{\citenamefont {Cros}(2014)}]{Cros2014}%
  \BibitemOpen
  \bibfield  {author} {\bibinfo {author} {\bibfnamefont {B.}~\bibnamefont {Cros}},\ }\bibfield  {title} {\bibinfo {title} {Laser-driven plasma wakefield: Propagation effects},\ }in\ \href {https://doi.org/10.5170/CERN-2016-001.207} {\emph {\bibinfo {booktitle} {CAS-CERN Accelerator School: Plasma Wake Acceleration 2014, Proceedings}}}\ (\bibinfo  {publisher} {CERN},\ \bibinfo {year} {2014})\ pp.\ \bibinfo {pages} {207--230}\BibitemShut {NoStop}%
\bibitem [{\citenamefont {Willingale}\ \emph {et~al.}(2013)\citenamefont {Willingale}, \citenamefont {Thomas}, \citenamefont {Nilson}, \citenamefont {Chen}, \citenamefont {Cobble}, \citenamefont {Craxton}, \citenamefont {Maksimchuk}, \citenamefont {Norreys}, \citenamefont {Sangster}, \citenamefont {Scott}, \citenamefont {Stoeckl}, \citenamefont {Zulick},\ and\ \citenamefont {Krushelnick}}]{Willingale_2013}%
  \BibitemOpen
  \bibfield  {author} {\bibinfo {author} {\bibfnamefont {L.}~\bibnamefont {Willingale}}, \bibinfo {author} {\bibfnamefont {A.~G.~R.}\ \bibnamefont {Thomas}}, \bibinfo {author} {\bibfnamefont {P.~M.}\ \bibnamefont {Nilson}}, \bibinfo {author} {\bibfnamefont {H.}~\bibnamefont {Chen}}, \bibinfo {author} {\bibfnamefont {J.}~\bibnamefont {Cobble}}, \bibinfo {author} {\bibfnamefont {R.~S.}\ \bibnamefont {Craxton}}, \bibinfo {author} {\bibfnamefont {A.}~\bibnamefont {Maksimchuk}}, \bibinfo {author} {\bibfnamefont {P.~A.}\ \bibnamefont {Norreys}}, \bibinfo {author} {\bibfnamefont {T.~C.}\ \bibnamefont {Sangster}}, \bibinfo {author} {\bibfnamefont {R.~H.~H.}\ \bibnamefont {Scott}}, \bibinfo {author} {\bibfnamefont {C.}~\bibnamefont {Stoeckl}}, \bibinfo {author} {\bibfnamefont {C.}~\bibnamefont {Zulick}},\ and\ \bibinfo {author} {\bibfnamefont {K.}~\bibnamefont {Krushelnick}},\ }\bibfield  {title} {\bibinfo {title} {Surface waves and electron acceleration from high-power, kilojoule-class laser interactions with
  underdense plasma},\ }\href {https://doi.org/10.1088/1367-2630/15/2/025023} {\bibfield  {journal} {\bibinfo  {journal} {New Journal of Physics}\ }\textbf {\bibinfo {volume} {15}},\ \bibinfo {pages} {025023} (\bibinfo {year} {2013})}\BibitemShut {NoStop}%
\bibitem [{\citenamefont {Snyder}\ \emph {et~al.}(2019)\citenamefont {Snyder}, \citenamefont {Ji}, \citenamefont {George}, \citenamefont {Willis}, \citenamefont {Cochran}, \citenamefont {Daskalova}, \citenamefont {Handler}, \citenamefont {Rubin}, \citenamefont {Poole}, \citenamefont {Nasir}, \citenamefont {Zingale}, \citenamefont {Chowdhury}, \citenamefont {Shen},\ and\ \citenamefont {Schumacher}}]{Snyder2019}%
  \BibitemOpen
  \bibfield  {author} {\bibinfo {author} {\bibfnamefont {J.}~\bibnamefont {Snyder}}, \bibinfo {author} {\bibfnamefont {L.~L.}\ \bibnamefont {Ji}}, \bibinfo {author} {\bibfnamefont {K.~M.}\ \bibnamefont {George}}, \bibinfo {author} {\bibfnamefont {C.}~\bibnamefont {Willis}}, \bibinfo {author} {\bibfnamefont {G.~E.}\ \bibnamefont {Cochran}}, \bibinfo {author} {\bibfnamefont {R.~L.}\ \bibnamefont {Daskalova}}, \bibinfo {author} {\bibfnamefont {A.}~\bibnamefont {Handler}}, \bibinfo {author} {\bibfnamefont {T.}~\bibnamefont {Rubin}}, \bibinfo {author} {\bibfnamefont {P.~L.}\ \bibnamefont {Poole}}, \bibinfo {author} {\bibfnamefont {D.}~\bibnamefont {Nasir}}, \bibinfo {author} {\bibfnamefont {A.}~\bibnamefont {Zingale}}, \bibinfo {author} {\bibfnamefont {E.}~\bibnamefont {Chowdhury}}, \bibinfo {author} {\bibfnamefont {B.~F.}\ \bibnamefont {Shen}},\ and\ \bibinfo {author} {\bibfnamefont {D.~W.}\ \bibnamefont {Schumacher}},\ }\bibfield  {title} {\bibinfo {title} {Relativistic laser driven electron accelerator using
  micro-channel plasma targets},\ }\href {https://doi.org/10.1063/1.5087409} {\bibfield  {journal} {\bibinfo  {journal} {Physics of Plasmas}\ }\textbf {\bibinfo {volume} {26}},\ \bibinfo {pages} {033110} (\bibinfo {year} {2019})}\BibitemShut {NoStop}%
\bibitem [{\citenamefont {Vranic}\ \emph {et~al.}(2018)\citenamefont {Vranic}, \citenamefont {Fonseca},\ and\ \citenamefont {Silva}}]{Vranic_2018}%
  \BibitemOpen
  \bibfield  {author} {\bibinfo {author} {\bibfnamefont {M.}~\bibnamefont {Vranic}}, \bibinfo {author} {\bibfnamefont {R.~A.}\ \bibnamefont {Fonseca}},\ and\ \bibinfo {author} {\bibfnamefont {L.~O.}\ \bibnamefont {Silva}},\ }\bibfield  {title} {\bibinfo {title} {Extremely intense laser-based electron acceleration in a plasma channel},\ }\href {https://doi.org/10.1088/1361-6587/aaa36c} {\bibfield  {journal} {\bibinfo  {journal} {Plasma Physics and Controlled Fusion}\ }\textbf {\bibinfo {volume} {60}},\ \bibinfo {pages} {034002} (\bibinfo {year} {2018})}\BibitemShut {NoStop}%
\bibitem [{\citenamefont {Sainte-Marie}\ \emph {et~al.}(2017)\citenamefont {Sainte-Marie}, \citenamefont {Gobert},\ and\ \citenamefont {Qu\'{e}r\'{e}}}]{Sainte-Marie2017}%
  \BibitemOpen
  \bibfield  {author} {\bibinfo {author} {\bibfnamefont {A.}~\bibnamefont {Sainte-Marie}}, \bibinfo {author} {\bibfnamefont {O.}~\bibnamefont {Gobert}},\ and\ \bibinfo {author} {\bibfnamefont {F.}~\bibnamefont {Qu\'{e}r\'{e}}},\ }\bibfield  {title} {\bibinfo {title} {Controlling the velocity of ultrashort light pulses in vacuum through spatio-temporal couplings},\ }\href {https://doi.org/10.1364/OPTICA.4.001298} {\bibfield  {journal} {\bibinfo  {journal} {Optica}\ }\textbf {\bibinfo {volume} {4}},\ \bibinfo {pages} {1298} (\bibinfo {year} {2017})}\BibitemShut {NoStop}%
\bibitem [{\citenamefont {Froula}\ \emph {et~al.}(2018)\citenamefont {Froula}, \citenamefont {Turnbull}, \citenamefont {Davies}, \citenamefont {Kessler}, \citenamefont {Haberberger}, \citenamefont {Palastro}, \citenamefont {Bahk}, \citenamefont {Begishev}, \citenamefont {Boni}, \citenamefont {Bucht}, \citenamefont {Katz},\ and\ \citenamefont {Shaw}}]{Froula2018}%
  \BibitemOpen
  \bibfield  {author} {\bibinfo {author} {\bibfnamefont {D.~H.}\ \bibnamefont {Froula}}, \bibinfo {author} {\bibfnamefont {D.}~\bibnamefont {Turnbull}}, \bibinfo {author} {\bibfnamefont {A.~S.}\ \bibnamefont {Davies}}, \bibinfo {author} {\bibfnamefont {T.~J.}\ \bibnamefont {Kessler}}, \bibinfo {author} {\bibfnamefont {D.}~\bibnamefont {Haberberger}}, \bibinfo {author} {\bibfnamefont {J.~P.}\ \bibnamefont {Palastro}}, \bibinfo {author} {\bibfnamefont {S.-W.}\ \bibnamefont {Bahk}}, \bibinfo {author} {\bibfnamefont {I.~A.}\ \bibnamefont {Begishev}}, \bibinfo {author} {\bibfnamefont {R.}~\bibnamefont {Boni}}, \bibinfo {author} {\bibfnamefont {S.}~\bibnamefont {Bucht}}, \bibinfo {author} {\bibfnamefont {J.}~\bibnamefont {Katz}},\ and\ \bibinfo {author} {\bibfnamefont {J.~L.}\ \bibnamefont {Shaw}},\ }\bibfield  {title} {\bibinfo {title} {Spatiotemporal control of laser intensity},\ }\href {https://doi.org/10.1038/s41566-018-0121-8} {\bibfield  {journal} {\bibinfo  {journal} {Nature Photonics}\ }\textbf {\bibinfo
  {volume} {12}},\ \bibinfo {pages} {262} (\bibinfo {year} {2018})}\BibitemShut {NoStop}%
\bibitem [{\citenamefont {Palastro}\ \emph {et~al.}(2020)\citenamefont {Palastro}, \citenamefont {Shaw}, \citenamefont {Franke}, \citenamefont {Ramsey}, \citenamefont {Simpson},\ and\ \citenamefont {Froula}}]{palastro2020dephasingless}%
  \BibitemOpen
  \bibfield  {author} {\bibinfo {author} {\bibfnamefont {J.}~\bibnamefont {Palastro}}, \bibinfo {author} {\bibfnamefont {J.}~\bibnamefont {Shaw}}, \bibinfo {author} {\bibfnamefont {P.}~\bibnamefont {Franke}}, \bibinfo {author} {\bibfnamefont {D.}~\bibnamefont {Ramsey}}, \bibinfo {author} {\bibfnamefont {T.}~\bibnamefont {Simpson}},\ and\ \bibinfo {author} {\bibfnamefont {D.}~\bibnamefont {Froula}},\ }\bibfield  {title} {\bibinfo {title} {Dephasingless laser wakefield acceleration},\ }\href@noop {} {\bibfield  {journal} {\bibinfo  {journal} {Physical review letters}\ }\textbf {\bibinfo {volume} {124}},\ \bibinfo {pages} {134802} (\bibinfo {year} {2020})}\BibitemShut {NoStop}%
\bibitem [{\citenamefont {Caizergues}\ \emph {et~al.}(2020)\citenamefont {Caizergues}, \citenamefont {Smartsev}, \citenamefont {Malka},\ and\ \citenamefont {Thaury}}]{Caizergues2020}%
  \BibitemOpen
  \bibfield  {author} {\bibinfo {author} {\bibfnamefont {C.}~\bibnamefont {Caizergues}}, \bibinfo {author} {\bibfnamefont {S.}~\bibnamefont {Smartsev}}, \bibinfo {author} {\bibfnamefont {V.}~\bibnamefont {Malka}},\ and\ \bibinfo {author} {\bibfnamefont {C.}~\bibnamefont {Thaury}},\ }\bibfield  {title} {\bibinfo {title} {Phase-locked laser-wakefield electron acceleration},\ }\href {https://doi.org/10.1038/s41566-020-0657-2} {\bibfield  {journal} {\bibinfo  {journal} {Nature Photonics}\ }\textbf {\bibinfo {volume} {14}},\ \bibinfo {pages} {475} (\bibinfo {year} {2020})}\BibitemShut {NoStop}%
\bibitem [{\citenamefont {Li}\ \emph {et~al.}(2024)\citenamefont {Li}, \citenamefont {Miller}, \citenamefont {Pierce}, \citenamefont {Mori}, \citenamefont {Thomas},\ and\ \citenamefont {Palastro}}]{li2024spatiotemporal}%
  \BibitemOpen
  \bibfield  {author} {\bibinfo {author} {\bibfnamefont {D.}~\bibnamefont {Li}}, \bibinfo {author} {\bibfnamefont {K.}~\bibnamefont {Miller}}, \bibinfo {author} {\bibfnamefont {J.}~\bibnamefont {Pierce}}, \bibinfo {author} {\bibfnamefont {W.}~\bibnamefont {Mori}}, \bibinfo {author} {\bibfnamefont {A.~G.}\ \bibnamefont {Thomas}},\ and\ \bibinfo {author} {\bibfnamefont {J.}~\bibnamefont {Palastro}},\ }\bibfield  {title} {\bibinfo {title} {Spatiotemporal control of high-intensity laser pulses with a plasma lens},\ }\href@noop {} {\bibfield  {journal} {\bibinfo  {journal} {Phys. Rev. Res.}\ }\textbf {\bibinfo {volume} {6}},\ \bibinfo {pages} {013272} (\bibinfo {year} {2024})}\BibitemShut {NoStop}%
\bibitem [{\citenamefont {Jolly}\ \emph {et~al.}(2020)\citenamefont {Jolly}, \citenamefont {Gobert}, \citenamefont {Jeandet},\ and\ \citenamefont {Qu{\'e}r{\'e}}}]{jolly2020controlling}%
  \BibitemOpen
  \bibfield  {author} {\bibinfo {author} {\bibfnamefont {S.~W.}\ \bibnamefont {Jolly}}, \bibinfo {author} {\bibfnamefont {O.}~\bibnamefont {Gobert}}, \bibinfo {author} {\bibfnamefont {A.}~\bibnamefont {Jeandet}},\ and\ \bibinfo {author} {\bibfnamefont {F.}~\bibnamefont {Qu{\'e}r{\'e}}},\ }\bibfield  {title} {\bibinfo {title} {Controlling the velocity of a femtosecond laser pulse using refractive lenses},\ }\href@noop {} {\bibfield  {journal} {\bibinfo  {journal} {Opt. Express}\ }\textbf {\bibinfo {volume} {28}},\ \bibinfo {pages} {4888} (\bibinfo {year} {2020})}\BibitemShut {NoStop}%
\bibitem [{\citenamefont {Kabacinski}\ \emph {et~al.}(2023)\citenamefont {Kabacinski}, \citenamefont {Oliva}, \citenamefont {Tissandier}, \citenamefont {Gautier}, \citenamefont {Kozlová}, \citenamefont {Goddet}, \citenamefont {Andriyash}, \citenamefont {Thaury}, \citenamefont {Zeitoun},\ and\ \citenamefont {Sebban}}]{Kabacinski2023}%
  \BibitemOpen
  \bibfield  {author} {\bibinfo {author} {\bibfnamefont {A.}~\bibnamefont {Kabacinski}}, \bibinfo {author} {\bibfnamefont {E.}~\bibnamefont {Oliva}}, \bibinfo {author} {\bibfnamefont {F.}~\bibnamefont {Tissandier}}, \bibinfo {author} {\bibfnamefont {J.}~\bibnamefont {Gautier}}, \bibinfo {author} {\bibfnamefont {M.}~\bibnamefont {Kozlová}}, \bibinfo {author} {\bibfnamefont {J.-P.}\ \bibnamefont {Goddet}}, \bibinfo {author} {\bibfnamefont {I.~A.}\ \bibnamefont {Andriyash}}, \bibinfo {author} {\bibfnamefont {C.}~\bibnamefont {Thaury}}, \bibinfo {author} {\bibfnamefont {P.}~\bibnamefont {Zeitoun}},\ and\ \bibinfo {author} {\bibfnamefont {S.}~\bibnamefont {Sebban}},\ }\bibfield  {title} {\bibinfo {title} {Spatio-temporal couplings for controlling group velocity in longitudinally pumped seeded soft x-ray lasers},\ }\href {https://doi.org/10.1038/s41566-023-01165-5} {\bibfield  {journal} {\bibinfo  {journal} {Nature Photonics}\ }\textbf {\bibinfo {volume} {17}},\ \bibinfo {pages} {354} (\bibinfo {year}
  {2023})}\BibitemShut {NoStop}%
\bibitem [{\citenamefont {Fu}\ \emph {et~al.}(2025)\citenamefont {Fu}, \citenamefont {Groussin}, \citenamefont {Liu}, \citenamefont {Mysyrowicz}, \citenamefont {Tikhonchuk},\ and\ \citenamefont {Houard}}]{fu2025steering}%
  \BibitemOpen
  \bibfield  {author} {\bibinfo {author} {\bibfnamefont {S.}~\bibnamefont {Fu}}, \bibinfo {author} {\bibfnamefont {B.}~\bibnamefont {Groussin}}, \bibinfo {author} {\bibfnamefont {Y.}~\bibnamefont {Liu}}, \bibinfo {author} {\bibfnamefont {A.}~\bibnamefont {Mysyrowicz}}, \bibinfo {author} {\bibfnamefont {V.}~\bibnamefont {Tikhonchuk}},\ and\ \bibinfo {author} {\bibfnamefont {A.}~\bibnamefont {Houard}},\ }\bibfield  {title} {\bibinfo {title} {Steering laser-produced \text{THz} radiation in air with superluminal ionization fronts},\ }\href@noop {} {\bibfield  {journal} {\bibinfo  {journal} {Phys. Rev. Lett.}\ }\textbf {\bibinfo {volume} {134}},\ \bibinfo {pages} {045001} (\bibinfo {year} {2025})}\BibitemShut {NoStop}%
\bibitem [{\citenamefont {Pigeon}\ \emph {et~al.}(2024)\citenamefont {Pigeon}, \citenamefont {Franke}, \citenamefont {Chong}, \citenamefont {Katz}, \citenamefont {Boni}, \citenamefont {Dorrer}, \citenamefont {Palastro},\ and\ \citenamefont {Froula}}]{Pigeon:24}%
  \BibitemOpen
  \bibfield  {author} {\bibinfo {author} {\bibfnamefont {J.~J.}\ \bibnamefont {Pigeon}}, \bibinfo {author} {\bibfnamefont {P.}~\bibnamefont {Franke}}, \bibinfo {author} {\bibfnamefont {M.~L.~P.}\ \bibnamefont {Chong}}, \bibinfo {author} {\bibfnamefont {J.}~\bibnamefont {Katz}}, \bibinfo {author} {\bibfnamefont {R.}~\bibnamefont {Boni}}, \bibinfo {author} {\bibfnamefont {C.}~\bibnamefont {Dorrer}}, \bibinfo {author} {\bibfnamefont {J.~P.}\ \bibnamefont {Palastro}},\ and\ \bibinfo {author} {\bibfnamefont {D.~H.}\ \bibnamefont {Froula}},\ }\bibfield  {title} {\bibinfo {title} {Ultrabroadband flying-focus using an axiparabola-echelon pair},\ }\href {https://doi.org/10.1364/OE.506112} {\bibfield  {journal} {\bibinfo  {journal} {Opt. Express}\ }\textbf {\bibinfo {volume} {32}},\ \bibinfo {pages} {576} (\bibinfo {year} {2024})}\BibitemShut {NoStop}%
\bibitem [{\citenamefont {Liberman}\ \emph {et~al.}(2024)\citenamefont {Liberman}, \citenamefont {Lahaye}, \citenamefont {Smartsev}, \citenamefont {Tata}, \citenamefont {Benracassa}, \citenamefont {Golovanov}, \citenamefont {Levine}, \citenamefont {Thaury},\ and\ \citenamefont {Malka}}]{Liberman2024}%
  \BibitemOpen
  \bibfield  {author} {\bibinfo {author} {\bibfnamefont {A.}~\bibnamefont {Liberman}}, \bibinfo {author} {\bibfnamefont {R.}~\bibnamefont {Lahaye}}, \bibinfo {author} {\bibfnamefont {S.}~\bibnamefont {Smartsev}}, \bibinfo {author} {\bibfnamefont {S.}~\bibnamefont {Tata}}, \bibinfo {author} {\bibfnamefont {S.}~\bibnamefont {Benracassa}}, \bibinfo {author} {\bibfnamefont {A.}~\bibnamefont {Golovanov}}, \bibinfo {author} {\bibfnamefont {E.}~\bibnamefont {Levine}}, \bibinfo {author} {\bibfnamefont {C.}~\bibnamefont {Thaury}},\ and\ \bibinfo {author} {\bibfnamefont {V.}~\bibnamefont {Malka}},\ }\bibfield  {title} {\bibinfo {title} {Use of spatiotemporal couplings and an axiparabola to control the velocity of peak intensity},\ }\href {https://doi.org/10.1364/OL.507713} {\bibfield  {journal} {\bibinfo  {journal} {Opt. Lett.}\ }\textbf {\bibinfo {volume} {49}},\ \bibinfo {pages} {814} (\bibinfo {year} {2024})}\BibitemShut {NoStop}%
\bibitem [{\citenamefont {Ramsey}\ \emph {et~al.}(2020)\citenamefont {Ramsey}, \citenamefont {Franke}, \citenamefont {Simpson}, \citenamefont {Froula},\ and\ \citenamefont {Palastro}}]{Ramsey2020}%
  \BibitemOpen
  \bibfield  {author} {\bibinfo {author} {\bibfnamefont {D.}~\bibnamefont {Ramsey}}, \bibinfo {author} {\bibfnamefont {P.}~\bibnamefont {Franke}}, \bibinfo {author} {\bibfnamefont {T.~T.}\ \bibnamefont {Simpson}}, \bibinfo {author} {\bibfnamefont {D.~H.}\ \bibnamefont {Froula}},\ and\ \bibinfo {author} {\bibfnamefont {J.~P.}\ \bibnamefont {Palastro}},\ }\bibfield  {title} {\bibinfo {title} {Vacuum acceleration of electrons in a dynamic laser pulse},\ }\href {https://doi.org/10.1103/PhysRevE.102.043207} {\bibfield  {journal} {\bibinfo  {journal} {Phys. Rev. E}\ }\textbf {\bibinfo {volume} {102}},\ \bibinfo {pages} {043207} (\bibinfo {year} {2020})}\BibitemShut {NoStop}%
\bibitem [{\citenamefont {Ramsey}\ \emph {et~al.}(2022)\citenamefont {Ramsey}, \citenamefont {Malaca}, \citenamefont {Di~Piazza}, \citenamefont {Formanek}, \citenamefont {Franke}, \citenamefont {Froula}, \citenamefont {Pardal}, \citenamefont {Simpson}, \citenamefont {Vieira}, \citenamefont {Weichman},\ and\ \citenamefont {Palastro}}]{Ramsey2022}%
  \BibitemOpen
  \bibfield  {author} {\bibinfo {author} {\bibfnamefont {D.}~\bibnamefont {Ramsey}}, \bibinfo {author} {\bibfnamefont {B.}~\bibnamefont {Malaca}}, \bibinfo {author} {\bibfnamefont {A.}~\bibnamefont {Di~Piazza}}, \bibinfo {author} {\bibfnamefont {M.}~\bibnamefont {Formanek}}, \bibinfo {author} {\bibfnamefont {P.}~\bibnamefont {Franke}}, \bibinfo {author} {\bibfnamefont {D.~H.}\ \bibnamefont {Froula}}, \bibinfo {author} {\bibfnamefont {M.}~\bibnamefont {Pardal}}, \bibinfo {author} {\bibfnamefont {T.~T.}\ \bibnamefont {Simpson}}, \bibinfo {author} {\bibfnamefont {J.}~\bibnamefont {Vieira}}, \bibinfo {author} {\bibfnamefont {K.}~\bibnamefont {Weichman}},\ and\ \bibinfo {author} {\bibfnamefont {J.~P.}\ \bibnamefont {Palastro}},\ }\bibfield  {title} {\bibinfo {title} {Nonlinear thomson scattering with ponderomotive control},\ }\href {https://doi.org/10.1103/PhysRevE.105.065201} {\bibfield  {journal} {\bibinfo  {journal} {Phys. Rev. E}\ }\textbf {\bibinfo {volume} {105}},\ \bibinfo {pages} {065201} (\bibinfo {year}
  {2022})}\BibitemShut {NoStop}%
\bibitem [{\citenamefont {Ye}\ \emph {et~al.}(2023)\citenamefont {Ye}, \citenamefont {Gu}, \citenamefont {Fan}, \citenamefont {Zhang}, \citenamefont {Wang}, \citenamefont {Tan}, \citenamefont {Zhang}, \citenamefont {Yang}, \citenamefont {Yan}, \citenamefont {Wen}, \citenamefont {Wu}, \citenamefont {Lu}, \citenamefont {Huang},\ and\ \citenamefont {Zhou}}]{Ye2023}%
  \BibitemOpen
  \bibfield  {author} {\bibinfo {author} {\bibfnamefont {H.}~\bibnamefont {Ye}}, \bibinfo {author} {\bibfnamefont {Y.}~\bibnamefont {Gu}}, \bibinfo {author} {\bibfnamefont {Q.}~\bibnamefont {Fan}}, \bibinfo {author} {\bibfnamefont {X.}~\bibnamefont {Zhang}}, \bibinfo {author} {\bibfnamefont {S.}~\bibnamefont {Wang}}, \bibinfo {author} {\bibfnamefont {F.}~\bibnamefont {Tan}}, \bibinfo {author} {\bibfnamefont {J.}~\bibnamefont {Zhang}}, \bibinfo {author} {\bibfnamefont {Y.}~\bibnamefont {Yang}}, \bibinfo {author} {\bibfnamefont {Y.}~\bibnamefont {Yan}}, \bibinfo {author} {\bibfnamefont {J.}~\bibnamefont {Wen}}, \bibinfo {author} {\bibfnamefont {Y.}~\bibnamefont {Wu}}, \bibinfo {author} {\bibfnamefont {W.}~\bibnamefont {Lu}}, \bibinfo {author} {\bibfnamefont {W.}~\bibnamefont {Huang}},\ and\ \bibinfo {author} {\bibfnamefont {W.}~\bibnamefont {Zhou}},\ }\bibfield  {title} {\bibinfo {title} {Enhanced thomson scattering x-ray sources with flying focus laser pulse},\ }\href {https://doi.org/10.1063/5.0130819}
  {\bibfield  {journal} {\bibinfo  {journal} {AIP Advances}\ }\textbf {\bibinfo {volume} {13}},\ \bibinfo {pages} {035330} (\bibinfo {year} {2023})}\BibitemShut {NoStop}%
\bibitem [{\citenamefont {Simpson}\ \emph {et~al.}(2024)\citenamefont {Simpson}, \citenamefont {Pigeon}, \citenamefont {Ambat}, \citenamefont {Miller}, \citenamefont {Ramsey}, \citenamefont {Weichman}, \citenamefont {Froula},\ and\ \citenamefont {Palastro}}]{simpson2024spatiotemporal}%
  \BibitemOpen
  \bibfield  {author} {\bibinfo {author} {\bibfnamefont {T.~T.}\ \bibnamefont {Simpson}}, \bibinfo {author} {\bibfnamefont {J.~J.}\ \bibnamefont {Pigeon}}, \bibinfo {author} {\bibfnamefont {M.~V.}\ \bibnamefont {Ambat}}, \bibinfo {author} {\bibfnamefont {K.~G.}\ \bibnamefont {Miller}}, \bibinfo {author} {\bibfnamefont {D.}~\bibnamefont {Ramsey}}, \bibinfo {author} {\bibfnamefont {K.}~\bibnamefont {Weichman}}, \bibinfo {author} {\bibfnamefont {D.~H.}\ \bibnamefont {Froula}},\ and\ \bibinfo {author} {\bibfnamefont {J.~P.}\ \bibnamefont {Palastro}},\ }\bibfield  {title} {\bibinfo {title} {Spatiotemporal control of two-color terahertz generation},\ }\href@noop {} {\bibfield  {journal} {\bibinfo  {journal} {Phys. Rev. Res.}\ }\textbf {\bibinfo {volume} {6}},\ \bibinfo {pages} {013041} (\bibinfo {year} {2024})}\BibitemShut {NoStop}%
\bibitem [{\citenamefont {Palastro}\ \emph {et~al.}(2018)\citenamefont {Palastro}, \citenamefont {Turnbull}, \citenamefont {Bahk}, \citenamefont {Follett}, \citenamefont {Shaw}, \citenamefont {Haberberger}, \citenamefont {Bromage},\ and\ \citenamefont {Froula}}]{Palastro2018}%
  \BibitemOpen
  \bibfield  {author} {\bibinfo {author} {\bibfnamefont {J.~P.}\ \bibnamefont {Palastro}}, \bibinfo {author} {\bibfnamefont {D.}~\bibnamefont {Turnbull}}, \bibinfo {author} {\bibfnamefont {S.-W.}\ \bibnamefont {Bahk}}, \bibinfo {author} {\bibfnamefont {R.~K.}\ \bibnamefont {Follett}}, \bibinfo {author} {\bibfnamefont {J.~L.}\ \bibnamefont {Shaw}}, \bibinfo {author} {\bibfnamefont {D.}~\bibnamefont {Haberberger}}, \bibinfo {author} {\bibfnamefont {J.}~\bibnamefont {Bromage}},\ and\ \bibinfo {author} {\bibfnamefont {D.~H.}\ \bibnamefont {Froula}},\ }\bibfield  {title} {\bibinfo {title} {Ionization waves of arbitrary velocity driven by a flying focus},\ }\href {https://doi.org/10.1103/PhysRevA.97.033835} {\bibfield  {journal} {\bibinfo  {journal} {Phys. Rev. A}\ }\textbf {\bibinfo {volume} {97}},\ \bibinfo {pages} {033835} (\bibinfo {year} {2018})}\BibitemShut {NoStop}%
\bibitem [{\citenamefont {Turnbull}\ \emph {et~al.}(2018)\citenamefont {Turnbull}, \citenamefont {Franke}, \citenamefont {Katz}, \citenamefont {Palastro}, \citenamefont {Begishev}, \citenamefont {Boni}, \citenamefont {Bromage}, \citenamefont {Milder}, \citenamefont {Shaw},\ and\ \citenamefont {Froula}}]{turnbull2018ionization}%
  \BibitemOpen
  \bibfield  {author} {\bibinfo {author} {\bibfnamefont {D.}~\bibnamefont {Turnbull}}, \bibinfo {author} {\bibfnamefont {P.}~\bibnamefont {Franke}}, \bibinfo {author} {\bibfnamefont {J.}~\bibnamefont {Katz}}, \bibinfo {author} {\bibfnamefont {J.}~\bibnamefont {Palastro}}, \bibinfo {author} {\bibfnamefont {I.}~\bibnamefont {Begishev}}, \bibinfo {author} {\bibfnamefont {R.}~\bibnamefont {Boni}}, \bibinfo {author} {\bibfnamefont {J.}~\bibnamefont {Bromage}}, \bibinfo {author} {\bibfnamefont {A.}~\bibnamefont {Milder}}, \bibinfo {author} {\bibfnamefont {J.}~\bibnamefont {Shaw}},\ and\ \bibinfo {author} {\bibfnamefont {D.}~\bibnamefont {Froula}},\ }\bibfield  {title} {\bibinfo {title} {Ionization waves of arbitrary velocity},\ }\href@noop {} {\bibfield  {journal} {\bibinfo  {journal} {Phys. Rev. Lett.}\ }\textbf {\bibinfo {volume} {120}},\ \bibinfo {pages} {225001} (\bibinfo {year} {2018})}\BibitemShut {NoStop}%
\bibitem [{\citenamefont {Wang}\ \emph {et~al.}(2020{\natexlab{a}})\citenamefont {Wang}, \citenamefont {Gong},\ and\ \citenamefont {Arefiev}}]{Wang2020_PoP}%
  \BibitemOpen
  \bibfield  {author} {\bibinfo {author} {\bibfnamefont {T.}~\bibnamefont {Wang}}, \bibinfo {author} {\bibfnamefont {Z.}~\bibnamefont {Gong}},\ and\ \bibinfo {author} {\bibfnamefont {A.}~\bibnamefont {Arefiev}},\ }\bibfield  {title} {\bibinfo {title} {Electron confinement by laser-driven azimuthal magnetic fields during direct laser acceleration},\ }\href {https://doi.org/10.1063/5.0006295} {\bibfield  {journal} {\bibinfo  {journal} {Physics of Plasmas}\ }\textbf {\bibinfo {volume} {27}},\ \bibinfo {pages} {053109} (\bibinfo {year} {2020}{\natexlab{a}})}\BibitemShut {NoStop}%
\bibitem [{\citenamefont {Gong}\ \emph {et~al.}(2021)\citenamefont {Gong}, \citenamefont {Hatsagortsyan},\ and\ \citenamefont {Keitel}}]{Gong2021}%
  \BibitemOpen
  \bibfield  {author} {\bibinfo {author} {\bibfnamefont {Z.}~\bibnamefont {Gong}}, \bibinfo {author} {\bibfnamefont {K.~Z.}\ \bibnamefont {Hatsagortsyan}},\ and\ \bibinfo {author} {\bibfnamefont {C.~H.}\ \bibnamefont {Keitel}},\ }\bibfield  {title} {\bibinfo {title} {Retrieving transient magnetic fields of ultrarelativistic laser plasma via ejected electron polarization},\ }\href {https://doi.org/10.1103/PhysRevLett.127.165002} {\bibfield  {journal} {\bibinfo  {journal} {Phys. Rev. Lett.}\ }\textbf {\bibinfo {volume} {127}},\ \bibinfo {pages} {165002} (\bibinfo {year} {2021})}\BibitemShut {NoStop}%
\bibitem [{\citenamefont {Cai}\ \emph {et~al.}(2025)\citenamefont {Cai}, \citenamefont {Gong}, \citenamefont {Shou}, \citenamefont {Wu}, \citenamefont {Ma}, \citenamefont {Qiu}, \citenamefont {Yang}, \citenamefont {Liu}, \citenamefont {Li}, \citenamefont {Zhou}, \citenamefont {Li}, \citenamefont {Xu},\ and\ \citenamefont {Yan}}]{Cai_2025}%
  \BibitemOpen
  \bibfield  {author} {\bibinfo {author} {\bibfnamefont {D.}~\bibnamefont {Cai}}, \bibinfo {author} {\bibfnamefont {Z.}~\bibnamefont {Gong}}, \bibinfo {author} {\bibfnamefont {Y.}~\bibnamefont {Shou}}, \bibinfo {author} {\bibfnamefont {X.}~\bibnamefont {Wu}}, \bibinfo {author} {\bibfnamefont {Q.}~\bibnamefont {Ma}}, \bibinfo {author} {\bibfnamefont {G.}~\bibnamefont {Qiu}}, \bibinfo {author} {\bibfnamefont {Z.}~\bibnamefont {Yang}}, \bibinfo {author} {\bibfnamefont {L.}~\bibnamefont {Liu}}, \bibinfo {author} {\bibfnamefont {J.}~\bibnamefont {Li}}, \bibinfo {author} {\bibfnamefont {Y.-F.}\ \bibnamefont {Zhou}}, \bibinfo {author} {\bibfnamefont {L.}~\bibnamefont {Li}}, \bibinfo {author} {\bibfnamefont {X.}~\bibnamefont {Xu}},\ and\ \bibinfo {author} {\bibfnamefont {X.}~\bibnamefont {Yan}},\ }\bibfield  {title} {\bibinfo {title} {Correlation between gamma photon emission and magnetic islands in laser-driven plasma channel},\ }\href {https://doi.org/10.1063/5.0270384} {\bibfield  {journal} {\bibinfo  {journal}
  {Physics of Plasmas}\ }\textbf {\bibinfo {volume} {32}},\ \bibinfo {pages} {073104} (\bibinfo {year} {2025})}\BibitemShut {NoStop}%
\bibitem [{\citenamefont {Rosmej}\ \emph {et~al.}(2021)\citenamefont {Rosmej}, \citenamefont {Shen}, \citenamefont {Pukhov}, \citenamefont {Antonelli}, \citenamefont {Barbato}, \citenamefont {Gyrdymov}, \citenamefont {Günther}, \citenamefont {Zähter}, \citenamefont {Popov}, \citenamefont {Borisenko},\ and\ \citenamefont {Andreev}}]{Rosmej2021}%
  \BibitemOpen
  \bibfield  {author} {\bibinfo {author} {\bibfnamefont {O.~N.}\ \bibnamefont {Rosmej}}, \bibinfo {author} {\bibfnamefont {X.~F.}\ \bibnamefont {Shen}}, \bibinfo {author} {\bibfnamefont {A.}~\bibnamefont {Pukhov}}, \bibinfo {author} {\bibfnamefont {L.}~\bibnamefont {Antonelli}}, \bibinfo {author} {\bibfnamefont {F.}~\bibnamefont {Barbato}}, \bibinfo {author} {\bibfnamefont {M.}~\bibnamefont {Gyrdymov}}, \bibinfo {author} {\bibfnamefont {M.~M.}\ \bibnamefont {Günther}}, \bibinfo {author} {\bibfnamefont {S.}~\bibnamefont {Zähter}}, \bibinfo {author} {\bibfnamefont {V.~S.}\ \bibnamefont {Popov}}, \bibinfo {author} {\bibfnamefont {N.~G.}\ \bibnamefont {Borisenko}},\ and\ \bibinfo {author} {\bibfnamefont {N.~E.}\ \bibnamefont {Andreev}},\ }\bibfield  {title} {\bibinfo {title} {Bright betatron radiation from direct-laser-accelerated electrons at moderate relativistic laser intensity},\ }\href {https://doi.org/10.1063/5.0042315} {\bibfield  {journal} {\bibinfo  {journal} {Matter and Radiation at Extremes}\
  }\textbf {\bibinfo {volume} {6}},\ \bibinfo {pages} {048401} (\bibinfo {year} {2021})}\BibitemShut {NoStop}%
\bibitem [{\citenamefont {Singh}\ \emph {et~al.}(2020)\citenamefont {Singh}, \citenamefont {Pathak}, \citenamefont {Shin}, \citenamefont {Choi}, \citenamefont {Nakajima}, \citenamefont {Lee}, \citenamefont {Sung}, \citenamefont {Lee}, \citenamefont {Rhee}, \citenamefont {Aniculaesei}, \citenamefont {Kim}, \citenamefont {Pae}, \citenamefont {Cho}, \citenamefont {Hojbota}, \citenamefont {Lee}, \citenamefont {Mollica}, \citenamefont {Malka}, \citenamefont {Ryu}, \citenamefont {Kim},\ and\ \citenamefont {Nam}}]{Singh2020}%
  \BibitemOpen
  \bibfield  {author} {\bibinfo {author} {\bibfnamefont {P.~K.}\ \bibnamefont {Singh}}, \bibinfo {author} {\bibfnamefont {V.~B.}\ \bibnamefont {Pathak}}, \bibinfo {author} {\bibfnamefont {J.~H.}\ \bibnamefont {Shin}}, \bibinfo {author} {\bibfnamefont {I.~W.}\ \bibnamefont {Choi}}, \bibinfo {author} {\bibfnamefont {K.}~\bibnamefont {Nakajima}}, \bibinfo {author} {\bibfnamefont {S.~K.}\ \bibnamefont {Lee}}, \bibinfo {author} {\bibfnamefont {J.~H.}\ \bibnamefont {Sung}}, \bibinfo {author} {\bibfnamefont {H.~W.}\ \bibnamefont {Lee}}, \bibinfo {author} {\bibfnamefont {Y.~J.}\ \bibnamefont {Rhee}}, \bibinfo {author} {\bibfnamefont {C.}~\bibnamefont {Aniculaesei}}, \bibinfo {author} {\bibfnamefont {C.~M.}\ \bibnamefont {Kim}}, \bibinfo {author} {\bibfnamefont {K.~H.}\ \bibnamefont {Pae}}, \bibinfo {author} {\bibfnamefont {M.~H.}\ \bibnamefont {Cho}}, \bibinfo {author} {\bibfnamefont {C.}~\bibnamefont {Hojbota}}, \bibinfo {author} {\bibfnamefont {S.~G.}\ \bibnamefont {Lee}}, \bibinfo {author} {\bibfnamefont
  {F.}~\bibnamefont {Mollica}}, \bibinfo {author} {\bibfnamefont {V.}~\bibnamefont {Malka}}, \bibinfo {author} {\bibfnamefont {C.~M.}\ \bibnamefont {Ryu}}, \bibinfo {author} {\bibfnamefont {H.~T.}\ \bibnamefont {Kim}},\ and\ \bibinfo {author} {\bibfnamefont {C.~H.}\ \bibnamefont {Nam}},\ }\bibfield  {title} {\bibinfo {title} {Electrostatic shock acceleration of ions in near-critical-density plasma driven by a femtosecond petawatt laser},\ }\bibfield  {journal} {\bibinfo  {journal} {Scientific Reports}\ }\textbf {\bibinfo {volume} {10}},\ \href {https://doi.org/10.1038/s41598-020-75455-1} {10.1038/s41598-020-75455-1} (\bibinfo {year} {2020})\BibitemShut {NoStop}%
\bibitem [{\citenamefont {Ospina-Bohórquez}\ \emph {et~al.}(2024)\citenamefont {Ospina-Bohórquez}, \citenamefont {Salgado-López}, \citenamefont {Ehret}, \citenamefont {Malko}, \citenamefont {Salvadori}, \citenamefont {Pisarczyk}, \citenamefont {Chodukowski}, \citenamefont {Rusiniak}, \citenamefont {Krupka}, \citenamefont {Guillon}, \citenamefont {Lendrin}, \citenamefont {Pérez-Callejo}, \citenamefont {Vlachos}, \citenamefont {Hannachi}, \citenamefont {Tarisien}, \citenamefont {Consoli}, \citenamefont {Verona}, \citenamefont {Prestopino}, \citenamefont {Dostal}, \citenamefont {Dudzak}, \citenamefont {Henares}, \citenamefont {Apiñaniz}, \citenamefont {Luis}, \citenamefont {Debayle}, \citenamefont {Caron}, \citenamefont {Ceccotti}, \citenamefont {Hernández-Martín}, \citenamefont {Hernández-Toro}, \citenamefont {Huault}, \citenamefont {Martín-López}, \citenamefont {Méndez}, \citenamefont {Nguyen-Bui}, \citenamefont {Perez-Hernández}, \citenamefont {Vaisseau}, \citenamefont {Varela}, \citenamefont
  {Volpe}, \citenamefont {Gremillet},\ and\ \citenamefont {Santos}}]{Ospina-Bohrquez2024}%
  \BibitemOpen
  \bibfield  {author} {\bibinfo {author} {\bibfnamefont {V.}~\bibnamefont {Ospina-Bohórquez}}, \bibinfo {author} {\bibfnamefont {C.}~\bibnamefont {Salgado-López}}, \bibinfo {author} {\bibfnamefont {M.}~\bibnamefont {Ehret}}, \bibinfo {author} {\bibfnamefont {S.}~\bibnamefont {Malko}}, \bibinfo {author} {\bibfnamefont {M.}~\bibnamefont {Salvadori}}, \bibinfo {author} {\bibfnamefont {T.}~\bibnamefont {Pisarczyk}}, \bibinfo {author} {\bibfnamefont {T.}~\bibnamefont {Chodukowski}}, \bibinfo {author} {\bibfnamefont {Z.}~\bibnamefont {Rusiniak}}, \bibinfo {author} {\bibfnamefont {M.}~\bibnamefont {Krupka}}, \bibinfo {author} {\bibfnamefont {P.}~\bibnamefont {Guillon}}, \bibinfo {author} {\bibfnamefont {M.}~\bibnamefont {Lendrin}}, \bibinfo {author} {\bibfnamefont {G.}~\bibnamefont {Pérez-Callejo}}, \bibinfo {author} {\bibfnamefont {C.}~\bibnamefont {Vlachos}}, \bibinfo {author} {\bibfnamefont {F.}~\bibnamefont {Hannachi}}, \bibinfo {author} {\bibfnamefont {M.}~\bibnamefont {Tarisien}}, \bibinfo {author}
  {\bibfnamefont {F.}~\bibnamefont {Consoli}}, \bibinfo {author} {\bibfnamefont {C.}~\bibnamefont {Verona}}, \bibinfo {author} {\bibfnamefont {G.}~\bibnamefont {Prestopino}}, \bibinfo {author} {\bibfnamefont {J.}~\bibnamefont {Dostal}}, \bibinfo {author} {\bibfnamefont {R.}~\bibnamefont {Dudzak}}, \bibinfo {author} {\bibfnamefont {J.~L.}\ \bibnamefont {Henares}}, \bibinfo {author} {\bibfnamefont {J.~I.}\ \bibnamefont {Apiñaniz}}, \bibinfo {author} {\bibfnamefont {D.~D.}\ \bibnamefont {Luis}}, \bibinfo {author} {\bibfnamefont {A.}~\bibnamefont {Debayle}}, \bibinfo {author} {\bibfnamefont {J.}~\bibnamefont {Caron}}, \bibinfo {author} {\bibfnamefont {T.}~\bibnamefont {Ceccotti}}, \bibinfo {author} {\bibfnamefont {R.}~\bibnamefont {Hernández-Martín}}, \bibinfo {author} {\bibfnamefont {J.}~\bibnamefont {Hernández-Toro}}, \bibinfo {author} {\bibfnamefont {M.}~\bibnamefont {Huault}}, \bibinfo {author} {\bibfnamefont {A.}~\bibnamefont {Martín-López}}, \bibinfo {author} {\bibfnamefont {C.}~\bibnamefont
  {Méndez}}, \bibinfo {author} {\bibfnamefont {T.~H.}\ \bibnamefont {Nguyen-Bui}}, \bibinfo {author} {\bibfnamefont {J.~A.}\ \bibnamefont {Perez-Hernández}}, \bibinfo {author} {\bibfnamefont {X.}~\bibnamefont {Vaisseau}}, \bibinfo {author} {\bibfnamefont {O.}~\bibnamefont {Varela}}, \bibinfo {author} {\bibfnamefont {L.}~\bibnamefont {Volpe}}, \bibinfo {author} {\bibfnamefont {L.}~\bibnamefont {Gremillet}},\ and\ \bibinfo {author} {\bibfnamefont {J.~J.}\ \bibnamefont {Santos}},\ }\bibfield  {title} {\bibinfo {title} {Laser-driven ion and electron acceleration from near-critical density gas targets: Towards high-repetition rate operation in the 1 pw, sub-100 fs laser interaction regime},\ }\bibfield  {journal} {\bibinfo  {journal} {Physical Review Research}\ }\textbf {\bibinfo {volume} {6}},\ \href {https://doi.org/10.1103/PhysRevResearch.6.023268} {10.1103/PhysRevResearch.6.023268} (\bibinfo {year} {2024})\BibitemShut {NoStop}%
\bibitem [{\citenamefont {Feder}\ \emph {et~al.}(2020)\citenamefont {Feder}, \citenamefont {Miao}, \citenamefont {Shrock}, \citenamefont {Goffin},\ and\ \citenamefont {Milchberg}}]{PhysRevResearch.2.043173}%
  \BibitemOpen
  \bibfield  {author} {\bibinfo {author} {\bibfnamefont {L.}~\bibnamefont {Feder}}, \bibinfo {author} {\bibfnamefont {B.}~\bibnamefont {Miao}}, \bibinfo {author} {\bibfnamefont {J.~E.}\ \bibnamefont {Shrock}}, \bibinfo {author} {\bibfnamefont {A.}~\bibnamefont {Goffin}},\ and\ \bibinfo {author} {\bibfnamefont {H.~M.}\ \bibnamefont {Milchberg}},\ }\bibfield  {title} {\bibinfo {title} {Self-waveguiding of relativistic laser pulses in neutral gas channels},\ }\href {https://doi.org/10.1103/PhysRevResearch.2.043173} {\bibfield  {journal} {\bibinfo  {journal} {Phys. Rev. Res.}\ }\textbf {\bibinfo {volume} {2}},\ \bibinfo {pages} {043173} (\bibinfo {year} {2020})}\BibitemShut {NoStop}%
\bibitem [{\citenamefont {Wang}\ \emph {et~al.}(2020{\natexlab{b}})\citenamefont {Wang}, \citenamefont {Ribeyre}, \citenamefont {Gong}, \citenamefont {Jansen}, \citenamefont {d'Humi\`eres}, \citenamefont {Stutman}, \citenamefont {Toncian},\ and\ \citenamefont {Arefiev}}]{Wang2020}%
  \BibitemOpen
  \bibfield  {author} {\bibinfo {author} {\bibfnamefont {T.}~\bibnamefont {Wang}}, \bibinfo {author} {\bibfnamefont {X.}~\bibnamefont {Ribeyre}}, \bibinfo {author} {\bibfnamefont {Z.}~\bibnamefont {Gong}}, \bibinfo {author} {\bibfnamefont {O.}~\bibnamefont {Jansen}}, \bibinfo {author} {\bibfnamefont {E.}~\bibnamefont {d'Humi\`eres}}, \bibinfo {author} {\bibfnamefont {D.}~\bibnamefont {Stutman}}, \bibinfo {author} {\bibfnamefont {T.}~\bibnamefont {Toncian}},\ and\ \bibinfo {author} {\bibfnamefont {A.}~\bibnamefont {Arefiev}},\ }\bibfield  {title} {\bibinfo {title} {Power scaling for collimated $\ensuremath{\gamma}$-ray beams generated by structured laser-irradiated targets and its application to two-photon pair production},\ }\href {https://doi.org/10.1103/PhysRevApplied.13.054024} {\bibfield  {journal} {\bibinfo  {journal} {Phys. Rev. Appl.}\ }\textbf {\bibinfo {volume} {13}},\ \bibinfo {pages} {054024} (\bibinfo {year} {2020}{\natexlab{b}})}\BibitemShut {NoStop}%
\bibitem [{\citenamefont {Arber}\ \emph {et~al.}(2015)\citenamefont {Arber}, \citenamefont {Bennett}, \citenamefont {Brady}, \citenamefont {Lawrence-Douglas}, \citenamefont {Ramsay}, \citenamefont {Sircombe}, \citenamefont {Gillies}, \citenamefont {Evans}, \citenamefont {Schmitz}, \citenamefont {Bell},\ and\ \citenamefont {Ridgers}}]{Arber2015}%
  \BibitemOpen
  \bibfield  {author} {\bibinfo {author} {\bibfnamefont {T.~D.}\ \bibnamefont {Arber}}, \bibinfo {author} {\bibfnamefont {K.}~\bibnamefont {Bennett}}, \bibinfo {author} {\bibfnamefont {C.~S.}\ \bibnamefont {Brady}}, \bibinfo {author} {\bibfnamefont {A.}~\bibnamefont {Lawrence-Douglas}}, \bibinfo {author} {\bibfnamefont {M.~G.}\ \bibnamefont {Ramsay}}, \bibinfo {author} {\bibfnamefont {N.~J.}\ \bibnamefont {Sircombe}}, \bibinfo {author} {\bibfnamefont {P.}~\bibnamefont {Gillies}}, \bibinfo {author} {\bibfnamefont {R.~G.}\ \bibnamefont {Evans}}, \bibinfo {author} {\bibfnamefont {H.}~\bibnamefont {Schmitz}}, \bibinfo {author} {\bibfnamefont {A.~R.}\ \bibnamefont {Bell}},\ and\ \bibinfo {author} {\bibfnamefont {C.~P.}\ \bibnamefont {Ridgers}},\ }\bibfield  {title} {\bibinfo {title} {Contemporary particle-in-cell approach to laser-plasma modelling},\ }\href {https://doi.org/10.1088/0741-3335/57/11/113001} {\bibfield  {journal} {\bibinfo  {journal} {Plasma Physics and Controlled Fusion}\ }\textbf {\bibinfo {volume}
  {57}},\ \bibinfo {pages} {113001} (\bibinfo {year} {2015})}\BibitemShut {NoStop}%
\bibitem [{\citenamefont {Franke}\ \emph {et~al.}(2021)\citenamefont {Franke}, \citenamefont {Ramsey}, \citenamefont {Simpson}, \citenamefont {Turnbull}, \citenamefont {Froula},\ and\ \citenamefont {Palastro}}]{Franke2021}%
  \BibitemOpen
  \bibfield  {author} {\bibinfo {author} {\bibfnamefont {P.}~\bibnamefont {Franke}}, \bibinfo {author} {\bibfnamefont {D.}~\bibnamefont {Ramsey}}, \bibinfo {author} {\bibfnamefont {T.~T.}\ \bibnamefont {Simpson}}, \bibinfo {author} {\bibfnamefont {D.}~\bibnamefont {Turnbull}}, \bibinfo {author} {\bibfnamefont {D.~H.}\ \bibnamefont {Froula}},\ and\ \bibinfo {author} {\bibfnamefont {J.~P.}\ \bibnamefont {Palastro}},\ }\bibfield  {title} {\bibinfo {title} {Optical shock enhanced self photon acceleration},\ }\href {https://doi.org/10.1103/PhysRevA.104.043520} {\bibfield  {journal} {\bibinfo  {journal} {Phys. Rev. A}\ }\textbf {\bibinfo {volume} {104}},\ \bibinfo {pages} {043520} (\bibinfo {year} {2021})}\BibitemShut {NoStop}%
\bibitem [{\citenamefont {Ridgers}\ \emph {et~al.}(2014)\citenamefont {Ridgers}, \citenamefont {Kirk}, \citenamefont {Duclous}, \citenamefont {Blackburn}, \citenamefont {Brady}, \citenamefont {Bennett}, \citenamefont {Arber},\ and\ \citenamefont {Bell}}]{Ridgers2014}%
  \BibitemOpen
  \bibfield  {author} {\bibinfo {author} {\bibfnamefont {C.~P.}\ \bibnamefont {Ridgers}}, \bibinfo {author} {\bibfnamefont {J.~G.}\ \bibnamefont {Kirk}}, \bibinfo {author} {\bibfnamefont {R.}~\bibnamefont {Duclous}}, \bibinfo {author} {\bibfnamefont {T.~G.}\ \bibnamefont {Blackburn}}, \bibinfo {author} {\bibfnamefont {C.~S.}\ \bibnamefont {Brady}}, \bibinfo {author} {\bibfnamefont {K.}~\bibnamefont {Bennett}}, \bibinfo {author} {\bibfnamefont {T.~D.}\ \bibnamefont {Arber}},\ and\ \bibinfo {author} {\bibfnamefont {A.~R.}\ \bibnamefont {Bell}},\ }\bibfield  {title} {\bibinfo {title} {Modelling gamma-ray photon emission and pair production in high-intensity laser-matter interactions},\ }\href {https://doi.org/10.1016/j.jcp.2013.12.007} {\bibfield  {journal} {\bibinfo  {journal} {Journal of Computational Physics}\ }\textbf {\bibinfo {volume} {260}},\ \bibinfo {pages} {273} (\bibinfo {year} {2014})}\BibitemShut {NoStop}%
\bibitem [{\citenamefont {Landau}\ and\ \citenamefont {Lifshitz}(1975)}]{landau_lifshitz_fields}%
  \BibitemOpen
  \bibfield  {author} {\bibinfo {author} {\bibfnamefont {L.~D.}\ \bibnamefont {Landau}}\ and\ \bibinfo {author} {\bibfnamefont {E.~M.}\ \bibnamefont {Lifshitz}},\ }\href@noop {} {\emph {\bibinfo {title} {The Classical Theory of Fields}}}\ (\bibinfo  {publisher} {Elsevier},\ \bibinfo {address} {Oxford},\ \bibinfo {year} {1975})\BibitemShut {NoStop}%
\end{thebibliography}%

\end{document}